\documentclass[a4paper,11pt]{article}
\pdfoutput=1 
\usepackage{graphicx}
\usepackage{float}
\usepackage{mciteplus}

\usepackage{jhep} 

\usepackage[T1]{fontenc} 

\renewcommand{\[}{\begin{equation}}
\renewcommand{\]}{\end{equation}}
\def\beq{\begin{equation}}
\def\eeq{\end{equation}}
\newcommand{\be}{\begin{eqnarray}}
\newcommand{\ee}{\end{eqnarray}}

\renewcommand{\texttt}{{}}

\usepackage{soul}

\def\bs{\begin{subequations}}
\def\es{\end{subequations}}

\def\Fc{\mathcal{F}}

\def\Hc{\mathcal{H}}

\def\Kc{\mathcal{K}}
\def\Lc{\mathcal{L}}
\def\Mc{\mathcal{M}}

\def\Oc{\mathcal{O}}
\def\Pc{\mathcal{P}}

\def\Rc{\mathcal{R}}

\def\Zc{\mathcal{Z}}

\newcommand{\tia}[1]{}

\newcommand{\bea}{\begin{eqnarray}}
\newcommand{\eea}{\end{eqnarray}}
\newcommand{\beas}{\begin{eqnarray*}}
\newcommand{\eeas}{\end{eqnarray*}}
\newcommand{\bal}{\begin{aligned}}
\newcommand{\eal}{\end{aligned}}

\def\({\left(}
\def\){\right)}

\newcommand{\LF}{\left(}
\newcommand{\RF}{\right)}
\newcommand{\LT}{\left[}
\newcommand{\RT}{\right]}

\newcommand{\intdg}[1]{\int d^{#1}x\sqrt{-g}}

\newcommand{\pd}{\partial}

\newcommand{\const}{\mathrm{const}}

\title{Non-Gaussianities and tensor-to-scalar ratio in non-local $R^{2}$-like inflation}

\author[a]{Alexey S. Koshelev,}

\author[b]{K. Sravan Kumar,}

\author[b]{Anupam Mazumdar,}

\author[c,d]{\\Alexei A. Starobinsky}

\affiliation[a~]{
	Departamento de F\'isica, Centro de Matem\'atica e Aplica\c{c}oes (CMA-UBI),
	Universidade da Beira Interior, 6200 Covilh\~a, Portugal }
\affiliation[b~]{
	Van Swinderen Institute, University of Groningen, 9747 AG Groningen,
	The Netherlands}
\affiliation[c~]{
	L. D. Landau Institute for Theoretical Physics RAS, Moscow 119334,
	Russian Federation}
\affiliation[d~]{Kazan Federal University, Kazan 420008, Republic of Tatarstan, Russian
	Federation}

\emailAdd{alexey@ubi.pt}
\emailAdd{sravan.korumilli@rug.nl,sravan.mph@gmail.com}
\emailAdd{anupam.mazumdar@rug.nl}
\emailAdd{alstar@landau.ac.ru}

\abstract{In this paper we will study $R^{2}$-like inflation
	in a non-local modification of gravity
	which contains quadratic in Ricci scalar and Weyl tensor terms with analytic infinite derivative form-factors in the action.
	It is known that the inflationary solution of the local $R+R^{2}$ 
	gravity remains a particular exact solution in this model. It 
	was shown earlier that the power spectrum of scalar perturbations generated
	during inflation in the non-local setup remains the same as in the local $R+R^2$
	inflation, whereas the power spectrum of tensor perturbations gets modified
	due to the non-local Weyl tensor squared term.
	In the present paper we go beyond 2-point correlators and compute the non-Gaussian parameter
	$f_{NL}$ related to $3$-point correlations generated during inflation,
	which we found to be different from those in the original local inflationary
	model and scenarios alike based on a local gravity. We evaluate non-local 
	corrections to the scalar bi-spectrum which give non-zero contributions to squeezed, 
	equilateral and orthogonal configurations. We show that $f_{NL}\sim {O}(1)$ with an arbitrary sign is achievable in this model based on the choice of form-factors and the scale of non-locality. We present the predictions for the  tensor-to-scalar ratio, $r$, and the tensor tilt, $n_t$. In contrast to standard inflation in a local gravity, here the possibility $n_t>0$ is not excluded. Thus, future CMB data can
	probe non-local behaviour of gravity at high space-time curvatures. }

\keywords{Models of Quantum Gravity, Cosmology of Theories beyond the SM}
\arxivnumber{arXiv:2003.00629}

\makeatother

\begin{document}
	
\maketitle

\section{Introduction}

After the Planck mission, inflationary cosmology took a turn towards the Starobinsky
inflationary model based on the modified $R+R^{2}$ gravity with one-loop corrections 
from matter quantum fields \cite{Starobinsky:1980te} (shortly dubbed as the (local) $R^2$ inflation 
afterwards), or to models with strongly non-minimally coupled scalar fields like the Higgs inflationary model 
\cite{Bezrukov:2007ep}, 
or to models emerged within string and supergravity (SUGRA) theories which have similar (very flat plateau-like) 
behaviour of their effective potentials in the Einstein frame \cite{Linde:2014nna}.
These highlighted models comprise a set of most successful ones from the point of view of the observational data matching but neither of them features a manifestly well established UV completion setting. 

In an attempt to tackle the issue of the UV completion for models of inflation on par with proposals to explain local higher order curvature terms like $R^{2}$ or non-minimal 
couplings of scalars by considering supergravity models \cite{Ketov:2010qz,Ketov:2012jt} there are recent developments on studying $R^{2}$ inflation solution within an analytic infinite derivative
(AID) gravity framework. This latter approach was shown to be absolutely consistent with existing cosmological observational data \cite{Biswas:2005qr,Biswas:2012bp,Briscese:2012ys,Craps:2014wga,Koshelev:2016xqb,Koshelev:2017tvv} while providing a viable track towards completing theory of quantum gravity \cite{Krasnikov:1987yj,Kuzmin:1989sp,Tomboulis:1997gg,Biswas:2011ar,Biswas:2013cha,Talaganis:2014ida,Modesto:2011kw,Tomboulis:2015esa,Talaganis:2016ovm,Dona:2015tra,Modesto:2014lga}. We therefore stick to this model in our effort to explore and analyze non-Gaussian perturbations and the tensor-to-scalar ratio during inflation in the present paper.

Analytic infinite derivative gravity essentially 
generalizes Einstein's General Relativity (GR) with quadratic in Ricci scalar, $R$ and Weyl tensor, $W$ terms in the action with analytic at zero functions of the d'Alembertian operator, also named form-factors, inserted in between like $R\Fc_R(\square)R$ where $\square$ is the covariant d'Alembertian operator, see: \cite{Biswas:2005qr,Biswas:2010zk,Biswas:2011ar,Biswas:2013kla,Biswas:2013cha,Koshelev:2013lfm,Biswas:2016etb,Biswas:2016egy}. Analyticity of form-factors at zero implies a smooth low-energy limit of the model. This type of gravity models is also often abbreviated as IDG for infinite derivative gravity. We however would mainly stick to the AID abbreviation to stress the analytic behavior of the form-factors for low momenta. This is an essential mathematical point and it distinguishes this type of gravity modifications from those having non-analytic dependence on derivatives, like it is already encountered in one-loop effective quantum gravity action \cite{Deser:1976yx,Starobinsky:1981zc,Barvinsky:1990up,Hawking:2000bb}, for example, non-analytic forms like logarithms of the d'Alembertian or inverse of the d'Alembertian appear in the effective action as a result of integrating out matter fields \cite{Deser:2007jk,Barvinsky:2014lja}.

Furthermore, it is the most general extension of GR around maximally symmetric space-times as it captures behavior of linear perturbations around such space-times in \textit{any} analytic gravity generalization (see \cite{Biswas:2016etb} for details).
{Inspiration for such a modification of gravity naturally comes from string field theory (SFT) where infinite derivative operators like $e^{{\square}/{\Mc_s^2}}$, with $\Mc_{s}$ being a string scale $\Mc_{s}\lesssim M_p$ enter the vertex terms of the string interaction \cite{Witten:1985cc,Witten:1986qs,Zwiebach:1985uq,Arefeva:2004odl,Ohmori:2001am,Calcagni:2005xc,Calcagni:2007ru}.} $M_p$ here is the Planck mass as usual. Discussing gravity we name $\Mc_s$ as the scale of non-locality.
Such a gravity is known to be ghost-free (unitary) improving thereby the proposal by Stelle \cite{Stelle:1976gc,Stelle:1977ry} which suffered from the presence of the spin-2 ghost. Absence of ghosts requires the presence of an infinite tower of derivatives and thus forces one to consider essentially non-local form-factors \cite{Biswas:2016egy}. This guaranties the power-counting renormalizability and
	raises the hope that Black-Hole and Big Bang singularities are being resolved in gravity theories of this kind.
{A number of crucial questions has been investigated recently regarding the resolution of singularities \cite{Biswas:2010zk,Conroy:2014dja,Conroy:2016sac,Edholm:2016hbt,Koshelev:2017bxd,Buoninfante:2018rlq,Buoninfante:2018stt,Buoninfante:2018gce,Buoninfante:2018mre,Buoninfante:2018xif,Buoninfante:2018xiw,Buoninfante:2019swn,Bosma:2019aiu,Buoninfante:2020ctr} in this approach}. 
The arguments above explain why such an infinite derivative gravity tends to become a UV complete gravity theory.

Obviously, the higher derivative form-factors provide the crux of the gravity modifications which we study in this paper. {We have freedom to choose form-factors such that the excitations spectrum is exactly like in a local $R^2$ gravity modification. Namely, there is a propagating massive scalar and a massless graviton in the theory. This scalar degree of freedom is identified with the `scalaron' that drives inflationary expansion and is responsible for nearly scale invariant fluctuations which in turn seed the large scale structure formation.} 
Although the structure of form-factors is constrained by the conditions on the theory to be ghost-free around maximally symmetric backgrounds \cite{Biswas:2016egy,Koshelev:2017ebj}, we lack for the time being a more fundamental approach of deriving them within the scope of string theory or SFT.  Worth mentioning that recently some progress has been made to fix form-factors within asymptotic safety approach to quantum gravity \cite{Bosma:2019aiu,Knorr:2019atm}. However, still observational cosmology is the only major way to get a guidance for any gravity modification and thus inflationary cosmology and CMB observations play a vital role.

It was shown in \cite{Koshelev:2017tvv} that an analytic non-local extension
of the $R^{2}$ inflation produces up to the leading order in the slow-roll expansion exactly the same value of the spectral index, $n_{s}$, of the Fourier power spectrum of primordial scalar perturbations generated during inflation, but the tensor to scalar power spectra ratio, $r$,  as well as the tensor power spectrum consistency relation, can get modified. Moreover, the modification of the tensor power spectrum is solely due to the form-factor in the quadratic in Weyl tensor term in the action.

In the present paper, we aim to understand and constrain the analytic infinite derivative form-factors in the considered gravity model using the current constraints on inflationary parameters \cite{Akrami:2018odb} such as $\LF n_s,\,r \RF$ and the non-Gaussianity parameter ($f_{NL}$) in the squeezed, equilateral and orthogonal configurations which read as 
\begin{equation}
\begin{aligned}
n_s= 0.9649\pm 0.0042\,\, \textrm{at}\,\, 68\% \textrm{CL}\,,\quad  r<0.064 \,\,\textrm{at}\,\, 95\%\, \textrm{CL}\,, 
\end{aligned}
\end{equation}
and 
\begin{equation}
f_{NL}^{sq}= -0.9\pm 5.1\,\quad f_{NL}^{equiv}= -26\pm 47\,\quad f_{NL}^{ortho}= -38\pm 24\,, \,\textrm{at}\,\, \,\,68\% \,\,\textrm{CL}. 
\label{lastest.fnl}
\end{equation}
with respect to Planck, BICEP2/Keck Array and BK15 with TT,TE, EE+lowE+lensing data \cite{Akrami:2018odb,Akrami:2019izv}. 

Non-Gaussianity is an important tool to understand primordial Universe and especially the nature of scalaron or inflaton\footnote{Which is a scalar degree of freedom coming from addition of hypothetical matter to GR.}, especially whether it is a canonical field, or a non-canonical one, or if multiple fields are responsible for the period of inflation, or if the inflation is driven effectively by a single field in the presence of relatively heavy fields, or if inflation happens in a higher order scalar tensor theories \cite{Maldacena:2002vr,Creminelli:2004yq,Chen:2006nt,Cheung:2007sv,DeFelice:2011zh,DeFelice:2013ar,Gao:2008dt,Arkani-Hamed:2015bza}. Every model produces a distinct signal for $f_{NL}$ for different triangular templates of 3-momenta. 
In the case of single canonical field models of inflation non-Gaussianities are very small as it was shown in \cite{Maldacena:2002vr}. {Local $R^2$ model can be seen as a single canonical field model upon conformal transformation of the $R+R^2$ gravity action to the Einstein frame where a minimally coupled scalar field with a very flat plateau shape potential appears \cite{DeFelice:2010aj}.} Any inflationary framework beyond a canonical scalar field may lead to different observable signatures of $f_{NL}$ \cite{Bartolo:2004if,Chen:2009zp}.
In our case there is only one canonical propagating scalar, namely the curvature perturbation, which is approximately time-independent on super-Hubble scales. However, the presence of higher derivative and even non-local operators can in principle lead to non-trivial signatures in the bi-spectrum (see for instance earlier observation of large non-Gaussianities in the context of SFT inspired $p$-adic inflation \cite{Barnaby:2006hi,Barnaby:2007yb} where a minimally coupled non-local scalar field drives inflation).

The paper is organized as follows. In Section~\ref{sec.2s} we present the modified gravity model to be studied and discuss the generic structure of form-factors for the theory to be ghost free. We summarize the general conditions on form-factor $\Fc_{R}\LF \square\RF$ for the theory to admit the local $R^2$ inflationary solution which satisfies the equation $\square R=M^2R$ where $M$ is the scalaron mass.
In Section~\ref{sec2a} we discuss second order inflationary perturbations and present quantitative results for scalar and tensor power spectra and their tilts for sample form-factors. In Appendix~\ref{sec:eom} we provide equations of motion, slow-roll conditions and further technicalities on second order scalar perturbations.
In Section~{\ref{sec.3rdac}} we compute the 3rd order variation of the studied gravity action around the local $R^2$ inflationary solution up to the leading order in the slow-roll approximation.
Using the mode functions computed in Section~\ref{sec.2s} we calculate the 3-point correlation functions and the non-linearity parameter $f_{NL}$ as functions of momenta.
Namely, we calculate the non-Gaussianity parameter $f_{NL}$ for local, equilateral and orthogonal configurations. To do quantitative analysis, we select sample form-factors and discuss the bounds on the scale of non-locality, $\Mc_{s}$, within the current CMB constraints. 
This section is supplemented by detailed computations in Appendices~\ref{app.use} and \ref{sec.3point}.
The Conclusion Section summarizes the performed analysis.

\section{Analytic infinite derivative gravity and inflationary solution}
\label{sec.2s}

In this section, we will review the equations of motion and inflationary solutions of AID gravity.
The notations commonly used throughout the paper are as follows. The reduced Planck mass in our notation is $\frac{M_{P}^{2}}{2}=\frac{1}{16\pi G}$ 
where $G$ is the Newtonian constant. The metric signature we work with is
$(-,+,+,+)$. 
The $4$-dimensional indices are labeled by small Greek letters and $3$-dimensional quantities are labeled by $i,j=1,2,3$. Throughout the paper a bar over quantities is used to indicate their background values for Friedmann-Lema\^itre-Robertson-Walker (FLRW) metric $g_{\mu\nu}=(-1,a^2,a^2,a^2)$, where $a=a(t)$ is the scale factor and $t$ is the cosmic time, like $\bar R$ for the background value of the Ricci scalar, $\bar \square$ for the background value of the d'Alembertian operator, etc. We use notation $^\prime$ and a dot to represent the differentiation with respect to conformal time $\tau$ and cosmic time $t$ respectively. $^{(\dagger)}$ and $^{(\ddagger)}$ denote first and second derivatives with respect to the argument for various functions. $^{(1,2,3)}$ denotes in most cases the order of the performed variation. $R, R_{\mu\nu},\, W_{\mu\nu\rho\sigma}$ denote the Ricci scalar, Ricci tensor and Weyl tensor respectively. Throughout the paper, the sign ``$\,\approx\,$'' denotes the leading order in the slow-roll approximation. 

The action we study contains non-local quadratic in curvature terms which was shown to be most general action around maximally symmetric space-times \cite{Biswas:2011ar,Biswas:2016egy,Biswas:2016etb,Koshelev:2016xqb,Koshelev:2017tvv}
\begin{equation}\label{NC-action}
S=\int d^{4}x\sqrt{-g}\left(\frac{M_{p}^{2}}{2}R+
	\frac{\lambda}{2}
\bigg[R\mathcal{F}_{R}\left(\square_{s}\right)R+W_{\mu\nu\rho\sigma}\mathcal{F}_{W}\left(\square_{s}\right)W^{\mu\nu\rho\sigma}\bigg]\right)\,.
\end{equation}
Here $\square_{s}={\square}/{\Mc_{s}^{2}}$ with $\Mc_{s}$ being the
scale of non-locality and $\lambda$ is a dimensionless parameter useful to control the effect of higher curvature contributions as the whole.
The form-factors are analytic at zero and can be Taylor expanded as follows
\begin{equation}
\Fc_{R}\LF \square_s \RF=\sum_{n=0}^\infty f_{Rn}\square_s^n,\quad \Fc_W\LF \square_s \RF= \sum_{n=0}^{\infty}f_{Wn}\square_s^n\,. 
\end{equation}
where $f_{Rn},f_{Wn}$ are dimensionless constants.
The vacuum equations of motion for the above action are given by (\ref{EoM}) in Appendix~\ref{sec:eom}.

The form-factors should not be arbitrary as they are restricted by the ghost-free conditions. For instance, the generic structure of form-factors that leads to an extra scalar field of mass $M$ and a massless tensor degree of freedom around maximally symmetric space-times backgrounds was studied in 
\cite{Biswas:2011ar,Biswas:2016etb,Biswas:2016egy}.

In the inflationary context we have to consider the structure of form-factors around the de Sitter background studied in \cite{Biswas:2016egy,Biswas:2016etb} which we discuss in the next Sections.
To study inflation, we start with 
FLRW metric which is conformally flat and as such the corresponding Weyl tensor vanishes.
Therefore the trace of equations (\ref{EoM}) for conformally flat backgrounds becomes 
\begin{equation}
E = (M_P^2-6\lambda\square\Fc_R(\square_s))R-\lambda(\Kc^\mu_\mu+2\tilde\Kc)=0\,,
\label{tEOMtrace}
\end{equation}
{where $\Kc^\mu_\mu,\, \tilde{\Kc}$ are infinite derivative square in scalar curvature terms defined in (\ref{KmnRFR-1}).}
In the sequel we put parameter $\lambda$ equal 1 for simplicity.
It was shown in \cite{Koshelev:2017tvv} that the local $R^2$ inflationary solution
	which satisfies the following equation
\begin{equation}
\bar{ \square}\bar{ R}=M^2\bar{ R}\,, \label{ansatz-1}\,
\end{equation}
	becomes the only (inflationary) solution\footnote{The local $R^2$ inflationary solution satisfies (\ref{ansatz-1}) with
	the scale factor
	\begin{equation}
	\begin{aligned}a & \approx a_{0}(t_{s}-t)^{-\frac{1}{6}}e^{-\frac{M^{2}(t_{s}-t)^{2}}{12}}\,,\quad 
	H=\frac{\dot{a}}{a} \approx\frac{M^{2}}{6}(t_{s}-t)+\frac{1}{6(t_{s}-t)}\,,\\
	\bar{R}=12H^{2}+6\dot{H} & \approx\frac{M^{4}(t_{s}-t)^{2}}{3}-\frac{M^{2}}{3}+{O}\left((t_{s}-t)^{-2}\right) \,,
	\end{aligned}
	\label{scale-fac}
	\end{equation}
	where $t_{s}\gg\frac{1}{M}$ denotes the end of inflation. }
	of AID gravity 
as long as form-factor $\Fc_{R}(\square_{s})$ obeys the following conditions 
\begin{equation}
\mathcal{F}_{R}^{\left(\dagger\right)}\left(\frac{M^{2}}{\Mc_{s}^{2}}\right)=0\quad,\quad\frac{M_{p}^{2}}{2}=3\lambda M^{2}\mathcal{F}_{1}\,,\label{FD1zero}
\end{equation}
where $\Fc_{R}^{\left(\dagger\right)}\left(\frac{M^{2}}{\Mc_{s}^{2}}\right) = 
\frac{\partial\Fc_{R}}{\partial \square_s}\Big\vert _{\square_s =\frac{M^2}{\Mc_{s}^2}}$ and $\Fc_{1} = \Fc_{R}\LF \frac{M^2}{\Mc_{s}^2} \RF$. Since cosmological 
observations indicate the nearly scale invariant power spectrum of scalar fluctuations, we consider the following hierarchy of scales\footnote{This can be seen 
heuristically by expanding the quadratic Ricci scalar part of the action (\ref{NC-action}) as 
\begin{equation}
S = \int d^4x\sqrt{-g} \LT \frac{M_p^2}{2}R+\frac{M_p^2}{12M^2}R^2+O \LF \frac{M_p^2 R\square R}{M^2\Mc_{s}^2}  \RF \RT\,.
\end{equation}
In the above action we see that at high curvature regime $R\gg M^2$ quadratic curvature terms are naturally dominant. However, it is easy to see that $R^2$ term is scale invariant (i.e., the term is invariant under local scale transformations $g_{\mu\nu}\to e^{\lambda}g_{\mu\nu}$ ) and all the higher derivative terms are not scale invariant. So it is obvious to see that the hierarchy $M^2\ll \Mc_s^2$ is essential for the model to be compatible with cosmological data.}
\cite{Koshelev:2016xqb,Koshelev:2017tvv}:
\[
M^{2}\ll \Mc_{s}^{2}\lesssim M_{p}^{2}\,,
\label{scale-hie}
\]
During inflation the following slow-roll condition
for the background solution (\ref{scale-fac}) is satisfied \cite{Starobinsky:1983zz,DeFelice:2010aj}:
\begin{equation}
\epsilon=-\frac{\dot{H}}{H^{2}}\,\approx\frac{M^{2}}{6H^{2}}\approx \frac{1}{2N}\ll1\implies M^{2}\ll H^{2}\,,\label{slw-roll}
\end{equation}
where $N\sim 50-60$ is number of e-foldings before the end of inflation.  

{Note that the Hubble-slow-roll parameters $\epsilon,\,\eta$ used here in the original Jordan frame differ from the potential slow-roll parameters $\epsilon_V,\,\eta_V$ usually known in the context of scalar field inflation (in the Einstein frame representation of the $R+R^2$ model) \cite{DeFelice:2010aj}. The second Hubble-slow-roll parameter in our case is $\eta = \epsilon+ \frac{d\ln\epsilon}{2dN} \approx \frac{1}{24N^2} \ll \epsilon$ (note that 
the calculation of its actual value requires using the second, next to leading, order of the slow-roll approximation presented in Eq. (\ref{scale-fac})). On the other hand, in the Einstein frame $\epsilon_V \ll |\eta_V|$. Since all our study is in Jordan frame, in all the computations we apply the slow-roll approximation up to the leading 
order in $\epsilon$. }

From the considerations (\ref{scale-hie})
and (\ref{slw-roll}) we assume the hierarchy of mass scales in the theory as shown on the scheme Fig.~1.
\begin{figure}[H]
\centering	\includegraphics[height=1in]{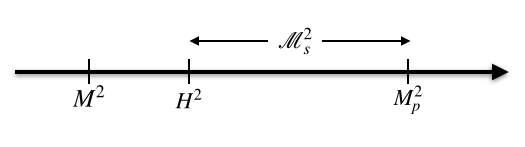}\caption{Hierarchy of mass scales in non-local $R^2$-like inflation.}
	\label{scales} 
\end{figure}
Further we refer to the condition (\ref{slw-roll}) as the slow-roll approximation.

\section{Second order perturbations and power spectra}
\label{sec2a}

Inflationary observables of $R^2$-like inflation in AID gravity related to two point correlations were studied in detail
in \cite{Craps:2014wga,Koshelev:2016xqb,Koshelev:2017tvv}. In this section, we will briefly review the second order perturbations of scalar and tensor parts and highlight the points essential for the rest of the paper without additional referencing.
We start with the following perturbed line element in terms of gauge invariant 
Bardeen potentials $\left(\Phi,\,\Psi\right)$ and transverse and traceless tensor fluctuation $h_{ij}$ 
\begin{equation}
ds^{2}=a^{2}\left(\tau\right)\left[-\left(1+2\Phi\right)d\tau^{2}+\left(\left(1-2\Psi\right)\delta_{ij}+2h_{ij}\right)dx^{i}dx^{j}\right]\,.\label{line-element}
\end{equation}
This is equivalent to fixing the Newtonian gauge.
Study of perturbed linear equations of motion in the slow-roll approximation, i.e. in the quasi-de Sitter regime gives among other important results
\begin{equation}
\left[\mathcal{F}_{1}\left(\bar{R}+3r_{1}\right)+\left(\bar{\square}-2H^{2}\right)\mathcal{F}_{W}\left(\bar{\square}+6H^{2}\right)\right]\frac{\Phi+\Psi}{a^{2}}  =0
\,.\label{phisizero}
\end{equation}
The solution of the above equation leads to $\Phi+\Psi\approx 0$ in the quasi-de Sitter limit in full analogy with the local $R^2$ inflationary model~\cite{Starobinsky:1981zc}. Proceeding with the construction of an action for a canonical scalar variable one must restrict the form-factor in order to avoid ghosts and this leads to the following form of the operator function $\Fc_R(\square_s)$:
\begin{equation}
\Fc_{R}\LF \square_s \RF = \Fc_{1}\frac{3e^{\gamma_S\LF \square_s \RF } \LF \square_s-\frac{M^2}{\Mc_{s}^2} \RF+\LF \frac{\bar{ R}}{\Mc_{s}^2}+3\frac{M^2}{\Mc_{s}^2} \RF}{3\square_s+\frac{\bar{ R}}{\Mc_{s}^2}} \,, 
\label{formfactors2s}
\end{equation}
where $\gamma_S$  is an entire function of the d'Alembertian operator.
Imposing the conditions (\ref{FD1zero}) on the form-factor $\Fc_R\LF\square_s\RF$ in (\ref{formfactors2s}),
we can deduce that
\begin{equation}
\gamma_S\LF \frac{M^2}{\Mc_{s}^2} \RF =0\,.
\label{gammas}
\end{equation}
We can also see that condition (\ref{gammas}) being used in (\ref{formfactors2s}) does not give us any relation between
scalaron mass and the non-locality scale. 

The second order action for the canonically normalized scalar has the form
\begin{equation}
	\delta^{(2)}S_{(S)}=\frac{1}{2}\int d\tau d^{3}x\sqrt{-\bar{g}}u\LF\bar{\square}-M^{2}\RF u\,,\label{canuac}
\end{equation}
where the subscript $(S)$ stands for the scalar part and $u \approx 2\Fc_{1}\bar{ R}\Psi$. A solution for Fourier modes of $\tilde{u}=au$ upon the spatial Fourier transform yields
in terms of Hankel functions
\cite{DeFelice:2010aj}
\begin{equation}
\tilde{u}_{k}=\frac{\sqrt{\pi}}{2}(-\tau)^{1/2}\LT c_{1k}H_{\nu_{s}}^{(1)}\LF-k\tau\RF+c_{2k}H_{\nu_{s}}^{(2)}\LF-k\tau\RF\RT\,.
\end{equation}

Adiabatic vacuum initial conditions for the observable range of Fourier modes of perturbations following from their quantization in the deep sub-Hubble
limit $\vert k\tau\vert\gg1$ are $c_{1k}=1,\,c_{2k}=0$. {The Bunch-Davies initial conditions taken literally would correspond to imposing these conditions for {\em all}  Fourier modes $\bf{k}$ including the limit $k\to 0$. However, this is not possible for the most realistic inflationary models including the local $R^2$ model. For our purposes, it is sufficient to impose the adiabatic vacuum initial conditions for all Fourier modes which are deep inside the Hubble radius at the beginning of the inflationary stage.} In the leading order in slow roll,
the sub-Hubble limit of the mode function can be approximated as
\begin{equation}
\tilde{u}_{k}=\frac{\Hc}{\sqrt{2k^{3}}}e^{ik\tau}(1-ik\tau)\,.\label{tildeuk}
\end{equation}

The curvature perturbation is defined as 
\begin{equation}
\Rc= \Psi+H\frac{\delta R}{\dot{\bar{R}}}=\Psi-\frac{24H^{3}}{24H\dot{H}}\Psi\approx-\frac{1}{\epsilon}\Psi=
 -\frac{1}{2\epsilon}\frac{1}{\sqrt{3\Fc_{1}\bar{R}}}\frac{H}{\sqrt{2k^3}}e^{ik\tau}\LF 1-ik\tau \RF\,. 
\label{curpert}
\end{equation}
and is (approximately) time-independent on super-Hubble scales $k\ll aH$. 
The primordial power spectrum and its tilt can be computed as
\begin{equation}
	\Pc_{\Rc}=\left.\frac{H^{2}}{16\pi^{2}\epsilon^{2}}\frac{1}{3\Fc_{1}\bar{R}}\right|_{k=aH}\,,\quad n_{s}\equiv\left.\frac{d\ln\Pc_{\Rc}}{d\ln k}\right|_{k=aH}\approx1-\frac{2}{N}\,,\label{Psuni}
\end{equation}
where $N$ is the number of $e$-folds and $\epsilon=-\frac{\dot{H}}{H^{2}}\approx\frac{1}{2N}$. We can notice that 
the scalar spectral index remains exactly same as for the local $R^2$ model. From the Planck data \cite{Ade:2015lrj} the power spectrum at the pivot scale $k_\ast = 0.05 \text{Mpc}^{-1}$ is $\Pc_\Rc^\ast\sim 2.2\times 10^{-9}$. Using this we get 
 the mass of scalaron, Hubble parameter and Ricci scalar at $k_\ast=aH$ as
 \begin{equation}
	 M\approx 1.3\times 10^{-5}\times \frac{55}{N_\ast}M_p,\quad H \approx \sqrt{\frac{M^2}{6\epsilon}} \approx 5.6\times 10^{-5}\times \frac{55}{N_\ast} M_p,\quad \bar{ R}_\ast \approx 220 M^2\times\frac{55}{N_\ast}\,. 
 \label{scalsn}
 \end{equation}
Here $N_\ast$ is the number of $e$-foldings before the end of inflation for pivot scale $k_\ast$.

The second order action for the canonically normalized tensor perturbations in the de Sitter approximation (applying $\bar{ R}\gg  M^2$) is  
\begin{equation}
\begin{aligned}\delta^{2}S_{(T)}= & \frac{1}{2}\int d^{4}x\sqrt{-\bar{g}}h_{ij}^{\perp} e^{2\gamma_T\LF \square_s-\frac{2\bar{ R}}{3\Mc_{s}^2} \RF}\LF\bar{\square}-\frac{\bar{R}}{6}\RF h^{\perp{ij}}\,,\end{aligned}
\label{tenspart}
\end{equation}
where the subscript $(T)$ stands for tensor part and $\gamma_T$ here is an entire function\footnote{Note that in the notation of \cite{Koshelev:2017tvv}, $\omega\LF \square_s \RF= \gamma_T\LF \square_s-\frac{2\bar{R}}{3\Mc_{s}^2} \RF$.} which is related to the form-factor as
\begin{equation}
\Fc_W\LF \square_s \RF = \frac{\Fc_{1}\bar{ R}}{\Mc_{s}^2}\frac{e^{\gamma_T
	\LF \square_s-\frac{2}{3}\frac{\bar{ R}}{\Mc_{s}^2} \RF}-1}{\square_s-\frac{2\bar{ R}}{3\Mc_{s}^2}}\,.
\label{fw}
\end{equation}

Following the standard procedure for getting the tensor power spectrum and its tilt one yields \cite{Koshelev:2017tvv}
\begin{equation}
\begin{aligned}
	\mathcal{P}_{\mathcal{T}} & =\left.\frac{1}{12\pi^{2}\mathcal{F}_{1}}\left(1-3\epsilon\right)e^{-2\gamma_T\left(-\frac{\bar{R}}{2\mathcal{M}_s^{2}}\right)}\right|_{k=aH}\,,\\
		n_{t}\equiv\left.\frac{d\ln\mathcal{P}_{\mathcal{T}}}{d\ln k}\right|_{k=aH} & \approx\left.-\frac{d\ln\mathcal{P}_{\mathcal{T}}}{dN}\left(1+\frac{1}{2N}\right)\right|_{k=aH}\\
	& \approx\left.-\frac{3}{2N^{2}}-\left(\frac{2}{N}+\frac{3}{2N^{2}}\right)\frac{\bar{R}}{2\mathcal{M}_s^{2}}\gamma_T^{(\dagger)}\left(-\frac{\bar{R}}{2\mathcal{M}_s^{2}}\right)\right|_{k=aH}\,,
\end{aligned}
\label{Ttilt}
\end{equation}
The crucial difference here in comparison with local $R^2$ model is that the tensor power spectrum is scaled by an exponential factor of $\gamma_T$ evaluated at the pole of the tensor mode $\bar{ \square}_s=\frac{\bar{ R}}{6\Mc_s^2}$ and accordingly tensor tilt also gets modified with a term proportional to the derivative of $\gamma_T$ at $\bar{ \square}_s=\frac{\bar{ R}}{6\Mc_s^2}$. 

The ratio of tensor-to-scalar power spectra is given by \cite{Koshelev:2017tvv}
\begin{equation}
	r=\left.\frac{12}{N^{2}}e^{-2\gamma_T\LF\frac{-\bar{R}}{2\Mc_s^{2}}\RF}\right|_{k=aH}\,.\label{T2S}
\end{equation}
In the local $R^2$ model $ r=\frac{12}{N^{2}}=3(1-n_s)^2$ as it follows from the original computation of scalar and tensor power spectra generated during inflation provided in \cite{Starobinsky:1983zz}. We can clearly notice that if $\gamma_T\LF\frac{-\bar{R}}{2\Mc_s^{2}}\RF=0$ one exactly recovers the same prediction as in the local $R^2$ model. A deviation from the local $R^2$ model happens if 
\begin{equation}
\gamma_T\LF\frac{-\bar{R}}{2\Mc_s^{2}}\RF \Bigg\vert_{k=aH}  \sim O\LF 1 \RF\,. 
\end{equation}
From (\ref{T2S}) and (\ref{Ttilt}) we can see that the values of $\LF r,\, n_t\RF$ explicitly depend 
on the ratio of scalar curvature to the square of $\Mc_{s}^2$ and the choice of entire function $\gamma_T\LF \square_s\RF$. The scalar curvature $\bar{ R}$ during inflation 
can be read from (\ref{scalsn}).  
The two observables $\LF r,\,n_t \RF$ can be fixed by the parameter space
\begin{equation}
\Bigg\{-\frac{\bar{ R}}{2\Mc_{s}^2},\, \gamma_T\LF - \frac{\bar{ R}}{2\Mc_{s}^2} \RF,\, \gamma_T^{(\dagger)}\LF -\frac{\bar{ R}}{2\Mc_{s}^2} \RF \Bigg\}\Bigg\vert_{k=aH}\,. 
\label{parame}
\end{equation}
From the latest Planck 2018 data \cite{Akrami:2018odb} we can deduce the following constraint (for $N_\ast=55$)
\begin{equation}
r< 0.064 \implies \gamma_T\LF -\frac{\bar{ R}}{2\Mc_{s}^2} \RF \Bigg\vert_{k=aH} > -1.32\,. 
\label{rcons}
\end{equation}
We have no constraint on the tensor tilt however. Imposing a heuristic limit on tensor tilt $-4\lesssim n_t \lesssim 4 $ we get a constraint on parameter space (\ref{parame}) as
\begin{equation}
-2N \lesssim -\frac{\bar{ R}}{2\Mc_{s}^2} \gamma_T^{(\dagger)}\LF -\frac{\bar{ R}}{2\Mc_{s}^2} \RF \Bigg\vert_{k=aH} \lesssim 2N\,. 
\label{ntcons}
\end{equation}

To illustrate novel configurations which become accessible thanks to the presence of the AID operators 
let us consider the following  form of the entire function
\begin{equation}
\gamma_T \LF  \square_s-\frac{2\bar{ R}}{3\Mc_{s}^2}  \RF \approx  \LF \square_{s}- \frac{2\bar{ R}}{3\Mc_{s}^2} \RF\LT \frac{\alpha_1}{3} \LF  \square_s-\frac{\bar{ R}_{\ast}}{6\Mc_{s}^2}\RF + \alpha_2 \LF  \square_s-\frac{2\bar{ R}}{3\Mc_{s}^2}\RF \RT \,. 
\label{fw1}
\end{equation}
where the approximation sign means that contributions of higher orders of the d'Alembertian operator are negligible when one evaluates this entire function and its derivative at $\square_s=\frac{\bar R}{6\Mc_s^2}$.
From the point of view of the UV completion the entire function here should provide a suppression of the propagator at large momenta and as such one should worry about the sign of the coefficient in front of the highest degree of the d'Alembertian operator \cite{Tomboulis:2015gfa}. On the other hand this implies that given there are terms $O(\square_s)$ in (\ref{fw1}) one can freely choose signs of parameters $\alpha_{1,2}$.

\begin{figure}[H]
	\includegraphics[width=75mm]{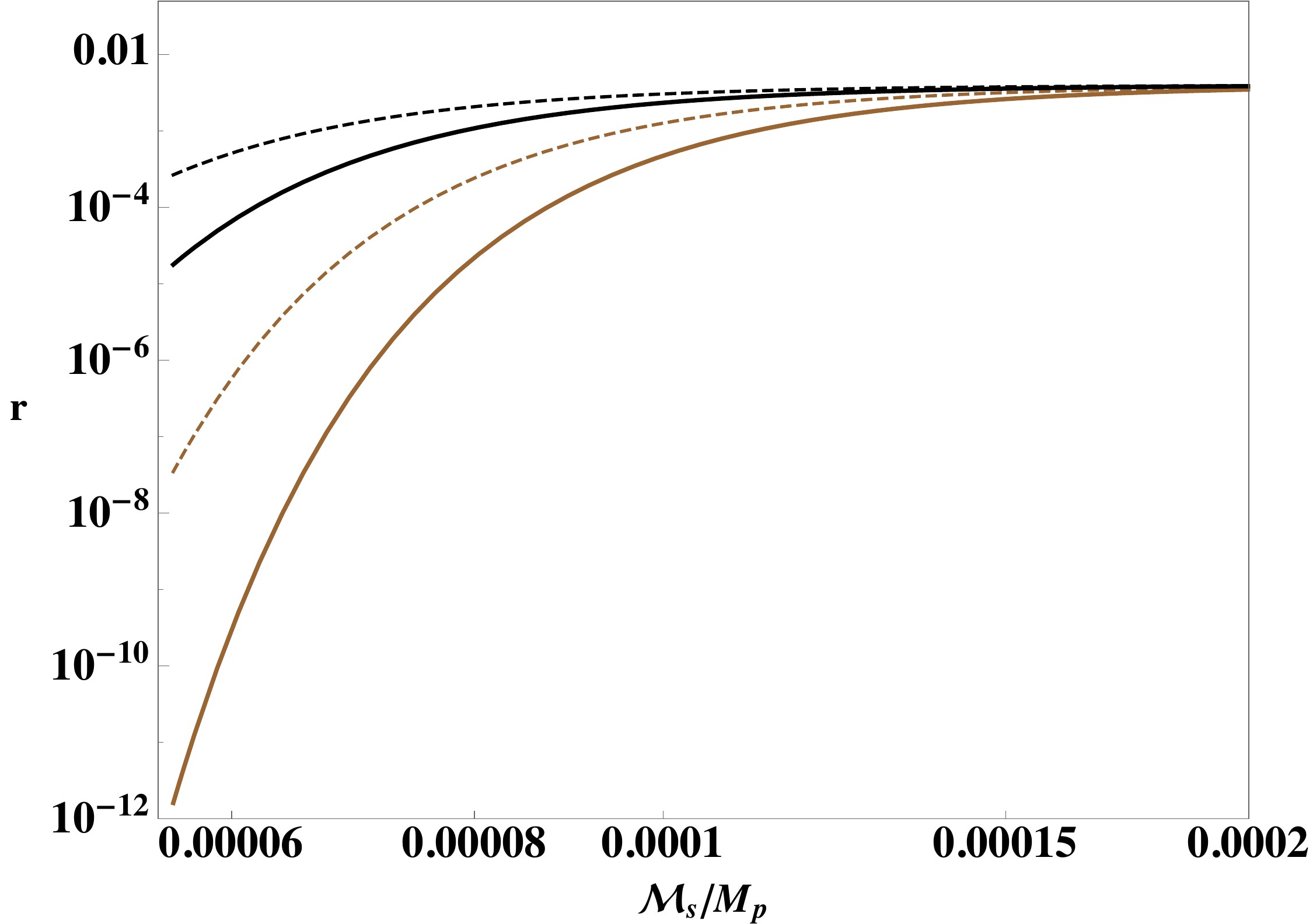}\quad \includegraphics[width=75mm]{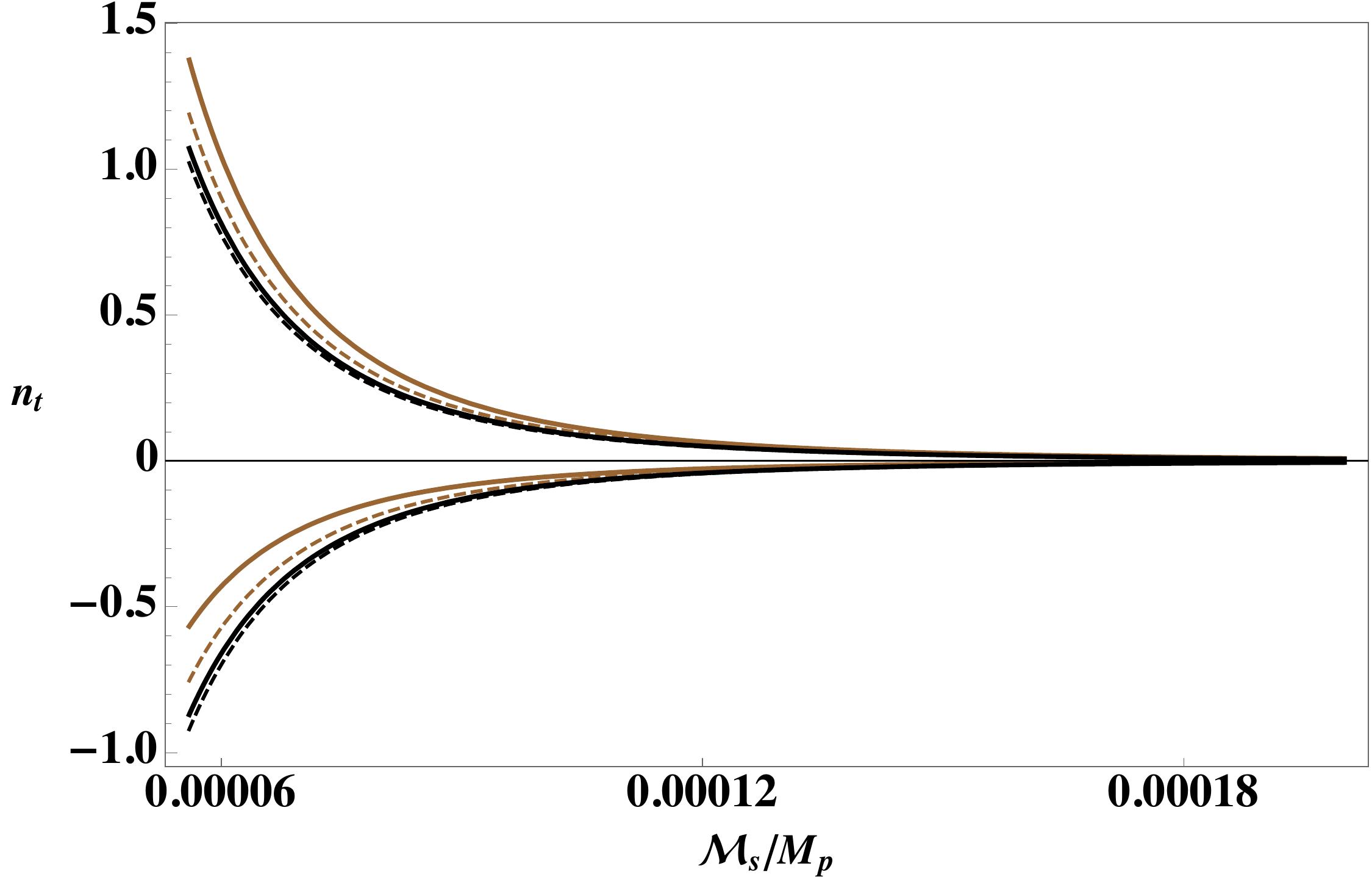}\caption{In the above plots, we depict $\LF r,\,n_t\RF$ versus the scale of non-locality $\Mc_{s}$ for the form-factor (\ref{fw1}) and $N_\ast=55$. 
	We have taken $\alpha_2= 0.52, 0.28, 0.13 , 0.065$ (corresponding to brown full and dashed, black full and dashed lines respectively).  In the right panel the lines with $n_t>0$ correspond to $\alpha_1= 2.5$ and the lines with $n_t<0$ correspond to $\alpha_1= - 2.5$. In both the plots, we can notice that in the limit $\Mc_{s}\to M_p$ we recover the predictions of the local $R^2$ model.}
	\label{trsnt} 
\end{figure}

\begin{figure}[H]
\includegraphics[width=70mm]{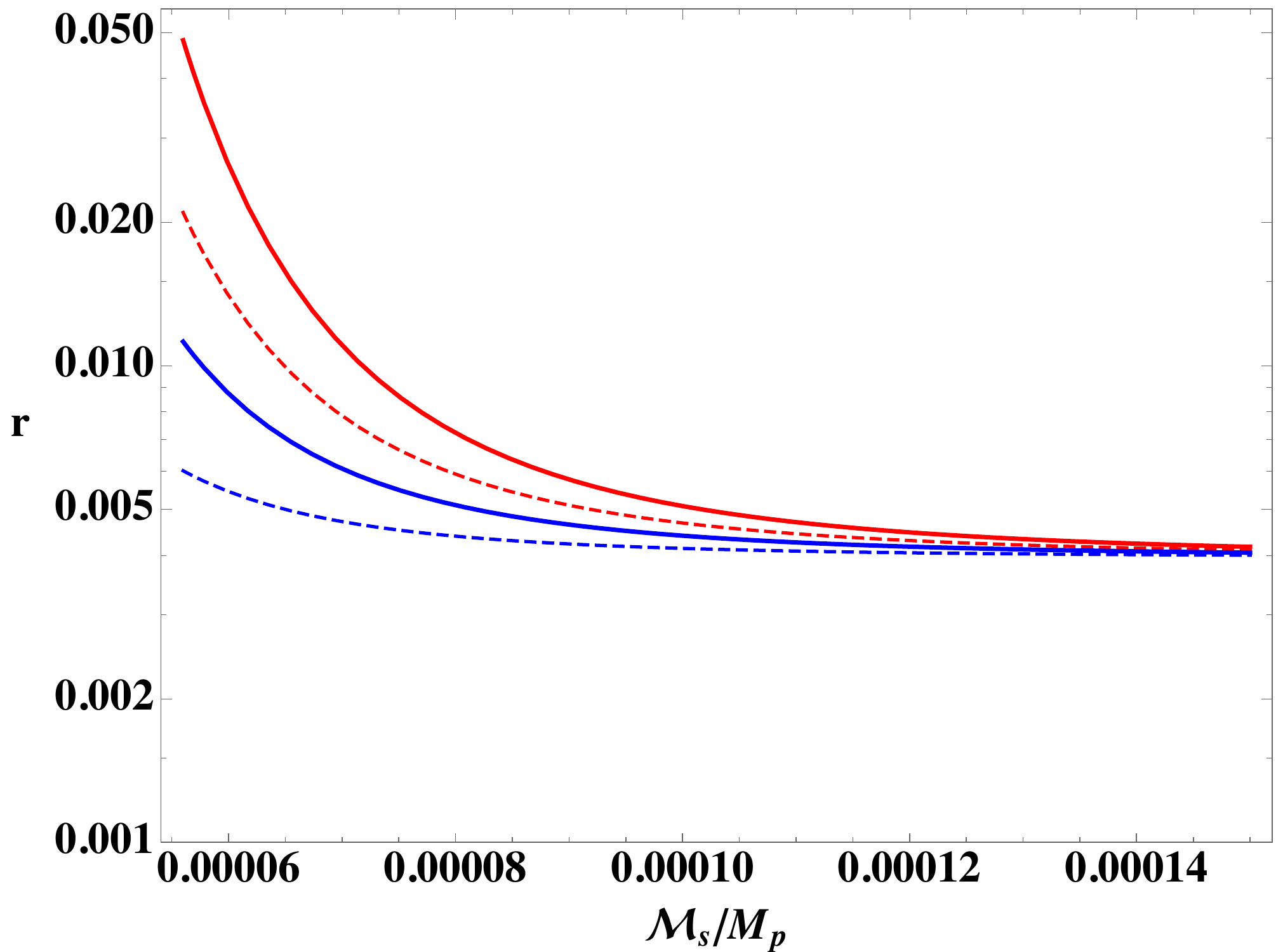}\quad \includegraphics[width=70mm]{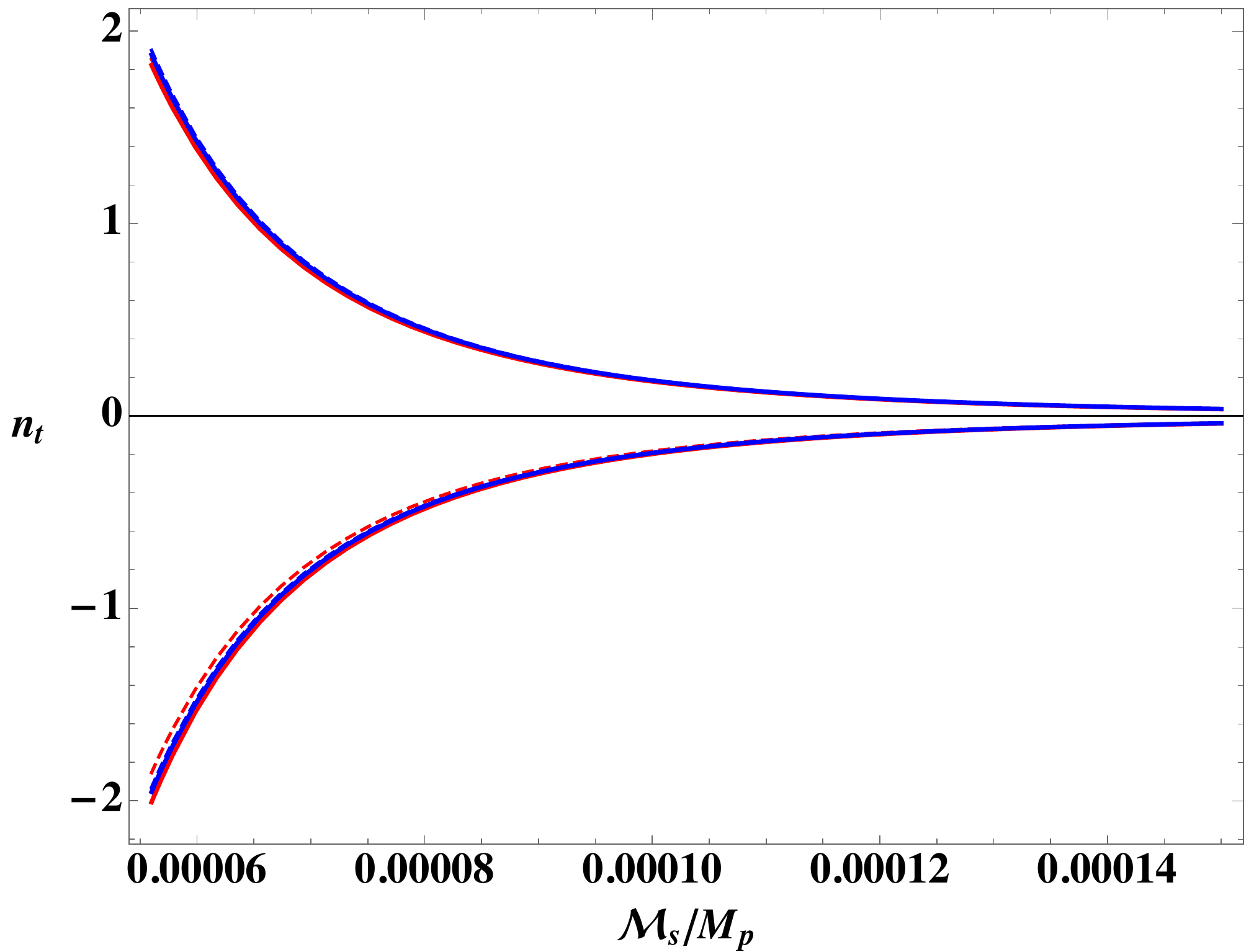}\caption{In the above plots, we depict $\LF r,\,n_t\RF$ versus the scale of non-locality $\Mc_{s}$ for the form-factor (\ref{fw1}) and $N_\ast=55$. 
	We have taken $\alpha_2= -0.06, -0.04, -0.025 , -0.01$ (corresponding to solid red, dashed red, solid blue and dashed blue lines respectively).  In the right panel the lines with $n_t>0$ correspond to $\alpha_1= 2.5$ and the lines with $n_t<0$ correspond to $\alpha_1= - 2.5$. In both the plots, we can notice that in the limit $\Mc_{s}\to M_p$ we recover the predictions of the local $R^2$ model. }
	\label{trsnt2} 
\end{figure}

In this particular example we see that $r$ depends on $\alpha_2$ and $\Mc_s$ because the coefficient of $\alpha_1$ vanishes when evaluated at $\square_s=\frac{\bar{ R}}{6\Mc_s^2}$. But the tensor tilt $n_t$ depends on $\LF\alpha_1,\,\alpha_2,\,\Mc_s\RF$. In figures Figs.~\ref{trsnt} and \ref{trsnt2} we present the predictions for $\LF r,\,n_t \RF$ for some values of parameter space $\LF\alpha_1,\,\alpha_2\RF$ and we take $\Mc_{s}\gtrsim H$ following the hierarchy in the scheme Fig.~\ref{scales}. 


\begin{figure}[H]
	\centering\includegraphics[width=150mm]{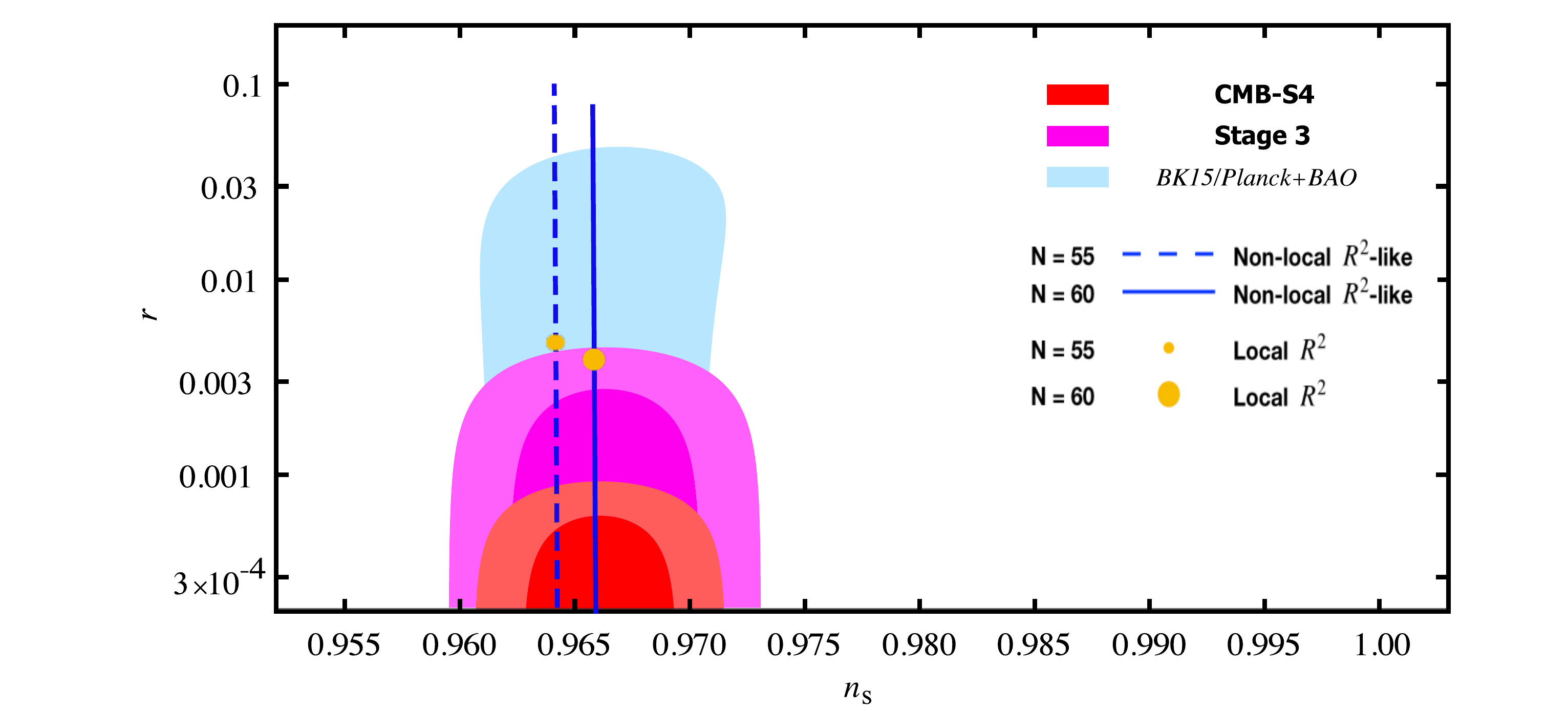}\caption{In the above plot, we depict the predictions of non-local $R^2$-like inflation in the $(n_s,r)$ plane of the latest CMB S-4 science paper about future forecast of detecting $B$-modes \cite{Abazajian:2019eic}.}
	\label{Staro-plot-1} 
\end{figure}

\begin{figure}[H]
	\centering\includegraphics[height=3in]{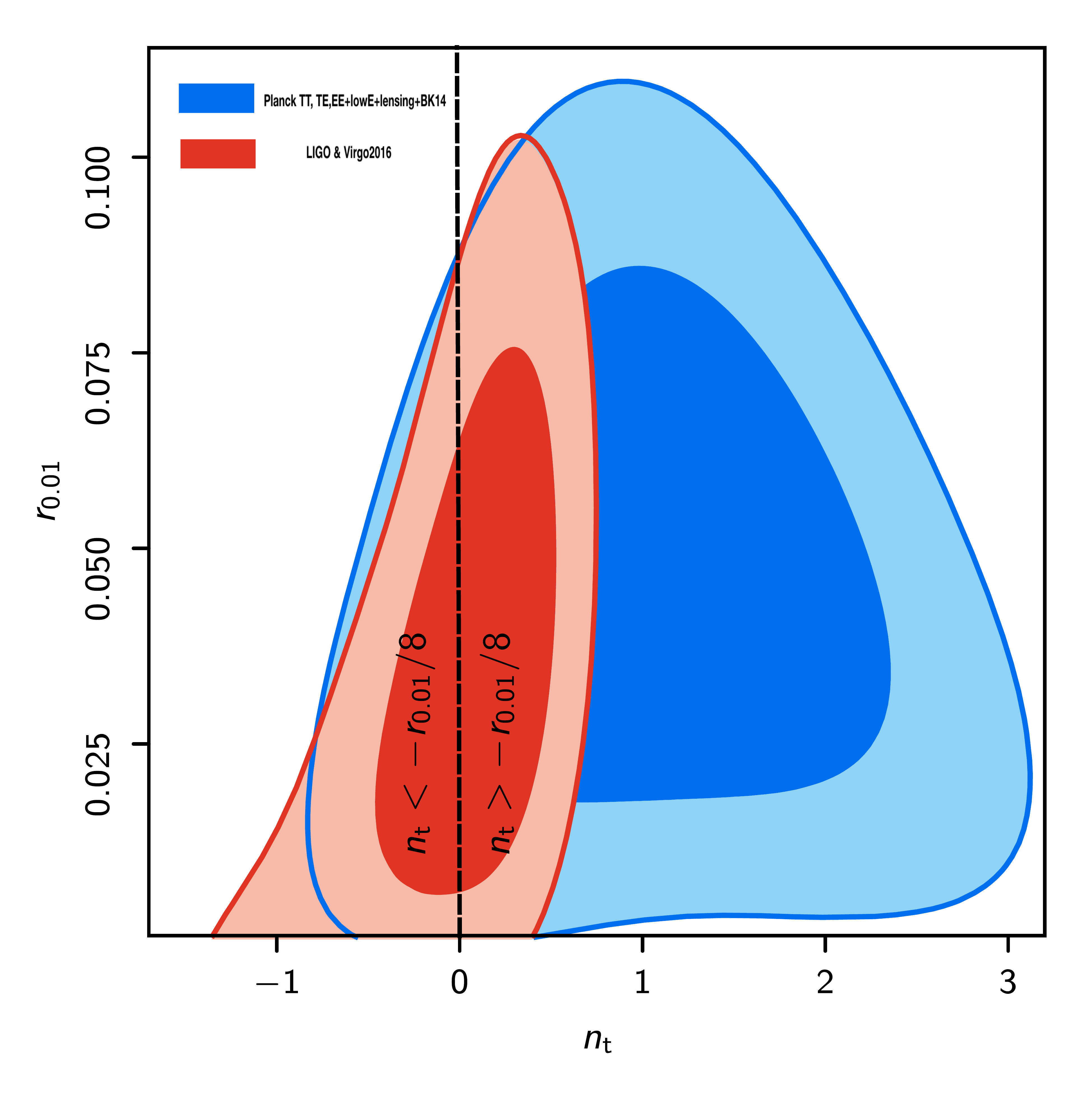}\caption{In the above plot, we note that the predictions of non-local $R^2$-like inflation can be anywhere within the likelihood projected $(n_t,r)$ plane of latest Planck 2018  \cite{Abazajian:2016yjj}.}
	\label{ntr} 
\end{figure}

Given we have free parameter space related to the form-factor between the Weyl tensors in the action, the predictions of the non-local $R^2$-like model for $
\LF r,\,n_s,\,n_t \RF$  lies within the future of scope of CMB experiments. 
Detection of primordial gravitational waves in the view of future CMB experiments such as BICEP2/Keck, CMB S-4, SO, LiteBIRD and PICO can fix the form-factor and the scale of non-locality \cite{Hui:2018cvg,Ade:2018gkx,Abazajian:2016yjj,Abazajian:2019eic,Ade:2018sbj,Hazumi:2019lys,Shandera:2019ufi,Hanany:2019lle}. In figure  Fig.~\ref{Staro-plot-1} we compare the predictions of non-local $R^2$-like model with the local $R^2$ model in $\LF n_s,\,r \RF$ plane.
Even though $r$ and $n_t$ are not detected so far, the latest Planck data presents a likelihood analysis for the values of $n_t$ which is an important parameter for inflationary setup as shown in figure Fig.~\ref{ntr}. In the single field models of inflation we get $r= -8n_t$ (which holds for the local $R^2$ model as well) and this is in general proposed to be a crucial test. In multifield models or non-canonical models of inflation this consistency relation gets violated due to isocurvature perturbations and/or non-zero sound speed of curvature perturbation \cite{Kobayashi:2013ina,Price:2014ufa}. In many of the inflationary setups (except some special cases \cite{Kobayashi:2011nu}) tensor tilt is found to be negative and it is regarded as a unique test for inflationary framework given that some alternative frameworks to the inflationary paradigm predict a blue tilt of tensor fluctuations \cite{Brandenberger:2015kga}.

As it becomes obvious in our case of a non-local $R^2$-like inflation we can achieve the blue tilt.
Therefore, we stress that a detection of a blue tensor tilt cannot rule
out inflation in general.

\section{Third order perturbations and inflationary bi-spectrum}
\label{sec.3rdac}

In this section, we compute the 3rd variation of the action (\ref{NC-action}) around (\ref{ansatz-1}) within the slow-roll approximation of the local $R^2$ inflationary solution. Then we compute the inflationary bi-spectrum of AID gravity and we especially quantify non-local corrections to the local $R^2$ gravity up to the leading order in the slow-roll approximation. 
Using the standard methods, the expectation value of 3-point correlations can be computed as \cite{Maldacena:2002vr,Seery:2005wm,DeFelice:2011zh,Khoury:2006fg,Weinberg:2005vy}

\begin{equation}
\langle\Rc\left(\mathbf{k_{1}}\right)\Rc\left(\mathbf{k_{2}}\right)\Rc\left(\mathbf{k_{3}}\right)\rangle  = -i \int_{-\infty}^{\tau_e}  a d\tau \langle 0\vert [ \Rc(\tau_{e},\,\mathbf{k_1})\Rc(\tau_{e},\,\mathbf{k_2})\Rc(\tau_{e},\,\mathbf{k_3}),\, H_{int} ] \vert 0 \rangle \,,
\label{3-point-f}
\end{equation}
where $\boldsymbol{k}_i$ are wave vectors and $H_{int}$ is the interaction Hamiltonian. It is related to a perturbation of the Lagrangian (\ref{NC-action}) expanded up to the 3rd order in curvature perturbations ($\Lc_3$) as $H_{int}\approx -\Lc_3$ that is a valid approximation during slow-roll inflationary regime \cite{Maldacena:2002vr,DeFelice:2011zh}. The integral (\ref{3-point-f}) describes non-linear evolution of inflationary perturbations produced in the adiabatic vacuum initial state which we compute in the limit when cosmological scales exit the Hubble radius. Since during quasi-de Sitter expansion $\tau\approx - \frac{1}{aH}$, it is a good approximation to calculate the integral in the limit 
$\tau_e\to 0$ \cite{DeFelice:2011zh}. In the Fourier space, the three-point correlation function of curvature perturbations is defined as
\begin{eqnarray}
\langle\Rc\left(\mathbf{k_{1}}\right)\Rc\left(\mathbf{k_{2}}\right)\Rc\left(\mathbf{k_{3}}\right)\rangle  =  \left(2\pi\right)^{3}\delta^{3}\left(\mathbf{k_{1}}+\mathbf{k_{2}}+\mathbf{k_{3}}\right)\mathcal{B}_{\Rc}\left(k_{1},k_{2},k_{3}\right)\,,
\end{eqnarray}
where $B_{\Rc}\left(k_{1},k_{2},k_{3}\right)$ is called 
the bi-spectrum and $\vert\boldsymbol{k_i}\vert=k_i$. Due to the translational invariance, the total momentum $\mathbf{K} = \mathbf{k_{1}}+\mathbf{k_{2}}+\mathbf{k_{3}}$ is conserved.
Scale invariance and rotational invariance imply that the 3-point function $B_\Rc\LF k_1,\,k_2,\,k_3 \RF$ has to be homogeneous function of degree $-6$ that means 
\begin{equation}
B_\Rc\LF \lambda k_1,\,\lambda k_2,\,\lambda k_3 \RF = \lambda^{-6}B_\Rc\LF k_1,\,k_2,\,k3 \RF\,.
\end{equation}
and the number of independent variables of the 3-point function reduces to 2 which are the ratios $\frac{k_2}{k_1},\,\frac{k_3}{k_1}$ \cite{Babich:2004gb}. 

In the template of CMB observations, non-linear corrections in the curvature perturbation expressed as \cite{Komatsu:2001rj,Takahashi:2014bxa}
\begin{equation}
\Rc = \Rc_g  -\frac{3}{5}f_{NL}\LF \Rc_g^2-\langle \Rc_g \rangle ^2 \RF \,, 
\end{equation} 
where $\Rc_g$ is the Gaussian random field and the $f_{NL}$ is the non-linearity parameter\footnote{The factor of $\frac{3}{5}$ comes from the relation between curvature perturbation $\Rc$ and the Bardeen variable $\Phi$ (Newtonian potential) during matter domination which reads as $\Rc= -\frac{5}{3}\Phi$,}. Then

\begin{equation}
B_{\Rc}\left(k_{1},k_{2},k_{3}\right) = 4\pi^4\frac{1}{\prod_i k_i^3} \Pc_{\Rc}^{ 2} A_{\Rc}\LF k_1,\,k_2,\,k_3 \RF \,, 
\end{equation}
where $A_\Rc\LF k_1,\,k_2,\,k_3 \RF$ is called the amplitude of bi-spectrum. In the momentum space, $f_{NL}$ is a function of three wave numbers and it is related to 
the amplitude of bi-spectrum as 
\begin{equation}
f_{NL} = -\frac{5}{6}\frac{A_\Rc\LF k_1,\,k_2,\,k_3 \RF}{\sum_i k_i^3}\,. 
\label{fnldef}
\end{equation}
The $f_{NL}$ defined above is often called as the reduced bi-spectrum. 

We calculate below the action (\ref{NC-action}) in the cubic order in curvature perturbations and the corresponding 3-point correlations (\ref{3-point-f})  up to the leading order in the slow-roll approximation using the following procedure\footnote{Usually, in scalar field models of inflation, the 3rd order action is computed using the Arnowitte-Deser-Misner (ADM) formalism where spatial slicing (reparametrizing spatial coordinates) is chosen such that perturbations of a scalar field always vanish. In this gauge choice the spatial part of the metric becomes $e^{2\Rc}\delta_{ij}dx^idx^j$ \cite{Seery:2005wm}. We however do computations in the Jordan frame and thus do not require such a gauge choice related to slicing of a 3-dimensional metric. We stick to a more convenient in our case gauge invariant approach which leads to the gauge invariant metric (\ref{line-element}). This corresponds to choosing the Newtonian gauge.}. 
We presented details of computations and results in the Appendix~\ref{sec.3point}.
\begin{itemize}
	\item We use the fact that $\Phi+\Psi\approx 0$ during the inflationary period as it was shown in \cite{Koshelev:2016xqb,Koshelev:2017tvv} and noted in the previous Section. 
	\item	From the second order action for canonical variable $\Upsilon$, we can deduce that 
	\begin{equation}
		\square_s\Psi\approx \frac{M^2}{\Mc_{s}^2}\Psi \implies \square_s^n\Psi \approx \LF\frac{M}{\Mc_{s}}\RF^{2n}\Psi\,. 
		\label{firstep}
	\end{equation}
	within the quasi-de Sitter approximation. In other words, we treat $\Psi$ as an eigenmode of the d'Alembertian in the computation of the 3rd order action up to the leading order 
slow-roll approximation. To perform this step, first we must eliminate the terms in the 3rd order action which are proportional to linearized equations of motion via a suitable field redefinition \cite{Maldacena:2002vr}.
	\item We carefully keep terms up to the  leading order in $\epsilon$ and neglect higher order contributions treating $\epsilon\approx \const$. In this regard, to convert the 3rd order action in $\Psi$ into curvature perturbation $\Rc$, we make substitution $\Psi \approx -\epsilon\Rc$ which follows from (\ref{curpert}). 
	\item Using the following approximations, we can bring the 3rd order action into a convenient form to compute 3-point correlations (\ref{3-point-f})  
	\begin{equation}
	\begin{aligned}
	\bar\square_s\Rc\approx &\, \frac{M^2}{\Mc_{s}^2}\Rc \implies & \Oc\LF\bar\square_s\RF \Rc \approx &\,\Oc\LF \frac{M^2}{\Mc_{s}^2} \RF\Rc \,\\
	\bar\square_s\Rc^{\prime}\approx &\, \LF\frac{\bar\square_s}{\Mc_s^2}+\frac{\bar{R}}{4\Mc_s^2}\RF \Rc^\prime \implies & \Oc\LF\bar\square_s\RF \Rc^\prime \approx &\, \Oc\LF \frac{M^2}{\Mc_{s}^2}+\frac{\bar{R}}{4\Mc_{s}^2} \RF\Rc^\prime\,, 
	\end{aligned}
	\end{equation}
	where $\Oc\LF\square_s\RF$ can be any analytic non-local operator which can be Taylor expanded in terms of d'Alembert operators. The second relation above is the 
result of the following general commutation relation \cite{SravanKumar:2019eqt} 
	\begin{equation}
	\nabla_{\mu}\square_s\phi=\square_s\nabla_{\mu}\phi-\frac{R_{\mu\nu}}{\Mc_{s}^2}\nabla^{\nu}\phi\,.
	\end{equation}
	where $\phi$ is a scalar. In the slow-roll approximation, $\bar{ R}_{\mu\nu}\approx\frac{\bar{ R}}{4}g_{\mu\nu}$, so
	we can approximate it as 
	\begin{equation}
	\begin{aligned}\nabla_{\mu}F(\square_s)\phi\approx &\, \Fc\LF\square_s-\frac{R}{4\Mc_{s}^2}\RF\nabla_{\mu}\phi\,.\end{aligned}
	\label{bpdc}
	\end{equation}
	
	\item We must be careful when applying the quasi-de Sitter approximation, especially when infinite derivatives are acting on a quantity. For example, the background Ricci scalar should not be treated as a constant until all the infinite tower of d'Alembertians are applied, otherwise we might end up with a zero contribution. As a consequence we would miss some terms when collecting next order non-zero contributions. For example, let us consider the following term
	\begin{equation}
	\begin{aligned}
		&\int d^4x\sqrt{-\bar{g}}\delta R\delta\Fc\LF\square_s\RF \delta R  \approx   4\epsilon^2\int d^4x\sqrt{-\bar{g}} \bar{R}\Rc\sum_{n=0}^{\infty}f_n\sum_{l=0}^{n-1}\bar\square_s^l\delta\square_s\bar\square_s^{n-l-1}\bar{R}\Rc\, \\ & \overset{\bar{R}=\const}{\approx}   4\epsilon^2\int d^4x\sqrt{-\bar{g}}\bar{R}\Rc\delta\square_s\Zc_{1}\LF \bar\square_s \RF\bar{R}\Rc\overset{\bar{R}=\const}{=}  0\,. 
	\end{aligned}
	\label{examplesl}
	\end{equation}
Looking carefully at the above derivation, we can see that passing from the first line to the second one, our assumption of approximating $\bar{R}\approx \const$ can be perfectly justified for every action of the d'Alembert operator, but using the same approximation in the next step would give us a null result due to the fact that $\Zc_{1}\LF \bar\square_s \RF\Rc = \Fc^{(\dagger)}_R\LF \frac{M^2}{\Mc_s^2} \RF\Rc = 0$ that follows from (\ref{FD1zero}).  Thus, we do not treat $\bar{R}=\const$ in the second line, rather we apply the background solution $\bar{ \square}\bar{R}=M^2\bar{R}$ and $\dot{\bar{R}}\approx -2\bar{R}H\epsilon$, and then we capture next order terms in the slow-roll approximation that provides us with new interactions of curvature perturbations and leads to a non-zero contribution to the 3-point function (c.f. (\ref{c4f})). The 
crucial lesson here is if any result of a computation gives zero, we must go to next to leading order in slow-roll, as far as we want to find a non-zero answer. 
	\item In our approach, perturbed modes which began in the adiabatic vacuum state when being deep inside the Hubble radius during inflation are evaluated when they left this region (rigorously speaking, after some amount of e-folds, large but still much less than $H^2/|\dot H|$, when they had reached their constant value in the super-Hubble regime).
	Given the fact the leading mode of curvature perturbations remains constant after that as far as $k\ll aH$, the three point correlations of $\Rc$ do not evolve there, too. This means that they keep information of primordial interactions of scalar modes during the period $-\infty < \tau< \tau_e$ where 
$\tau_e\sim -\frac{1}{K}$ is a (conformal) time after few e-foldings of the Hubble radius exit \cite{Pajer:2016ieg}.
\end{itemize}	
Using the above steps, we write below the result of a long and tedious computation presented in the Appendix~\ref{sec.3point}. Our obtained cubic order action in 
$\Rc$ of AID gravity in the leading order slow-roll approximation is 
\begin{equation}
\begin{aligned}
	\delta^{(3)}S_{(S)} =  & 4\epsilon M_p^2 \int   d\tau d^3x \Bigg\{T_1 \Rc\nabla\Rc\cdot\nabla\Rc+
	T_2 \Rc\Rc^{\prime 2}+ T_3 \Hc^2\Rc^3\\+&
T_4	\Hc \Rc\Rc\Rc^\prime+ T_5\Hc^{-1}\nabla\Rc\cdot\nabla\Rc\Rc^\prime + T_6 \Hc^{-1}\Rc^{\prime 3}+ T_7 \Hc^{-2}\Rc^\prime \nabla\Rc\cdot\nabla\Rc^\prime  \Bigg\}\,, 
\end{aligned}
\label{3rda}
\end{equation}
where $T_i$'s are dimensionless constants which can be read from (\ref{Tterms}). Here we present the final result for the amplitude of bi-spectrum after numerous simplifications and neglecting terms that are higher order in slow-roll:

	\begin{equation}
	\begin{aligned}
	A_\Rc\LF k_1,\,k_2,\,k_3 \RF = &\, -\LF 2\epsilon+\frac{3\epsilon^2}{4} \RF A_1+\LF 2\epsilon+\frac{3\epsilon^2}{4} \RF A_2-\frac{\epsilon^2}{2} A_3\\
	&+ \frac{\bar{ R}}{M_p^2}\epsilon^3 T_{\textrm{NL}} \Bigg[ \frac{16}{3}\epsilon A_2-32A_4+2A_5-2A_6+\frac{16}{3}\epsilon A_7 \Bigg]\\ &+ \frac{4\bar{ R}^2}{M_p^2\Mc_{s}^2}\epsilon^4\Fc^{(\dagger)}_R\LF \frac{\bar{ R}}{4\Mc_{s}^2} \RF 
\Bigg( A_2+\frac{1}{3}A_7 \Bigg) \,, 
	\end{aligned}
	\label{finalfNL}
	\end{equation}
	where the terms $A_i$ indicate the contributions from 8 types of interactions of curvature perturbations which can be read from (\ref{A1}) to (\ref{A8}) 
		\begin{equation}
	\begin{aligned}
	A_1 =  & 2\boldsymbol{k}_1\cdot \boldsymbol{k}_2 \LT K-\frac{k_1k_2+k_2k_3+k_3k_1}{K} -\frac{k_1k_2k_3}{K^2} \RT +\textrm{perms}\, ,\\
	A_2 = & \frac{2k_1^2k_2^2}{K} +\frac{2k_1^2k_2^2k_3}{K^2}+\textrm{perms}\,, \\
	A_3 = & - \frac{K^3}{3} +2Kk_1k_2+\frac{k_1k_2k_3}{3}  +\text{perms}\,\\
	A_4 = & k_3^2\LT -\frac{4K}{3}-\frac{2k_1k_2}{K} -\frac{2k_1+2k_2}{3} \RT +\text{perms}\,,\\
	A_5 = & 2\LF \boldsymbol{k}_1\cdot \boldsymbol{k}_2 \RF k_3^2\LT \frac{2}{K}+\frac{2k_1+2k_2}{K^2}+\frac{4k_1k_2}{K^3} \RT +\text{perms}\,, \\
	A_6 = & \frac{12k_1^2k_2^2k_3^2}{K^3}\,, \\ 
	A_7= &  \LF \boldsymbol{k}_2\cdot \boldsymbol{k}_3 \RF k_1^2k_3^2\LT -\frac{2}{K^3}-\frac{6k_2}{K^4} \RT +\text{perms}\,,\\ 
	\end{aligned}
	\end{equation}
where $K=k_1+k_2+k_3$ and
	\begin{equation}
   \begin{aligned}
	   T_{\textrm{NL}} & = \Bigg[ \Fc_{R}\LF \frac{M^2}{\Mc_{s}^2} +\frac{\bar{ R}}{4\Mc_{s}^2}\RF - \Fc_{1} \Bigg]  \Bigg\vert _{K=aH} \\ &\approx \Bigg[ \Fc_{R}\LF \frac{\bar{ R}}{4\Mc_{s}^2}\RF - \Fc_{1} +\epsilon\frac{\bar{ R}}{8\Mc_{s}^2}\Fc^{(\dagger)}_R\LF \frac{\bar{ R}}{4\Mc_{s}^2}\RF \Bigg]\Bigg \vert  _{K=aH}\,. 
\end{aligned}
\label{Tterms1}
	\end{equation}
The non-Gaussianity parameter $f_{NL}$ can be obtained by substituting (\ref{finalfNL}) in ({\ref{fnldef}}). 

In the AID gravity, we can qualitatively see that we depart from the local theory due to the additional non-local interactions of the curvature perturbation listed from (\ref{A4}) to (\ref{A9}). This can for example imply that the long wavelength mode that is exited the Hubble radius can strongly interact with the short wavelength mode that is crossing the Hubble radius. Although this effect is present in the case of standard single field inflation \cite{Maldacena:2002vr,Chen:2010xka}, it is significantly suppressed by slow-roll. Enhanced non-Gaussianities beyond the standard single field only known to occur in the context of non-canonical and/or multifield models of inflation \cite{Chen:2010xka,Byrnes:2014pja}. Our case significantly differs from this so far well known physics of inflation. We have obtained enhanced non-Gaussianities due to non-localities present in AID gravity leading to non-local interactions of the only propagating scalar mode of the theory\footnote{Note that the sound speed of curvature perturbation in our case is unity, so our result is fundamentally very different from single field non-canonical models \cite{Chen:2010xka}.}. 
A trivial observation we can make is that the contributions (\ref{A4}) to (\ref{A9})  do not appear in local theory. From the observational point of view, three configurations of reduced bi-spectrum $f_{NL}$ for squeezed $(k_1\to0, k_2= k_3=\frac{k}{2})$, equilateral $(k_1=k_2=k_3=k)$ and orthogonal $(k_1=k_2=k/4, k_3=k/2)$ are the most relevant.
	
	 We can easily verify that our computation of bi-spectrum reduces to the result of the local $R^2$ model in the limit $T_{\textrm{NL}} \to 0 $ and $\Fc_{R}^{(\dagger)}\LF \frac{\bar{R}}{4\Mc_{s}^2}\RF \to 0$. Especially, we can verify here the result of squeezed shape ($k_1\ll k_2\sim k_3 $)
	\begin{equation}
	f_{NL} \approx  \frac{5}{12}\LF 1-n_s \RF +O\LF \epsilon^2 \RF\,, 
	\label{fnlloc}
	\end{equation}
	which is exactly the Maldacena consistency relation found in the single canonical scalar field inflation \cite{Maldacena:2002vr}. {Note here that the Maldacena consistency relation was alternatively derived on kinematic grounds and was proven to be valid in general for single clock slow-roll models of inflation \cite{Creminelli:2004yq,Cheung:2007sv}. The proof is based on the fact that in the standard slow-roll inflation the long wavelength mode that is far outside of the horizon effectively implies a renormalization of the background field and the corresponding rescaling of the short wavelength mode \cite{Creminelli:2004yq,Seery:2005wm}. However, it is known that even in the case of a single field slow-roll
		inflation, the consistency relation can be violated if the second, time
		dependent, mode of curvature perturbations ${\cal R}$ (often dubbed as the
		decaying mode) becomes important temporarily, so that the total ${\cal R}$
		can not be considered as a constant during this period~\cite{Ganc:2010ff,
			Namjoo:2012aa,Martin:2012pe}. In this connection, there is a very instructive calculation
			of the bi-spectrum for the single field inflationary
			model~\cite{Starobinsky:2005ab} that has $n_s=1$ {\em exactly}, i.e.
			outside the slow-roll approximation. In this case, contrary to the zero non-Gaussianity parameter value predicted
			by the Maldacena consistency relation, $f_{NL}$ is non-zero and strongly
			($\propto k_1^2$) scale-dependent \cite{Ganc:2010ff}. This example shows also
			that the violation of the consistency condition for the bi-spectrum does
			not necessarily lead to the appearance of any new features in the spectrum
			itself.}
		
			{In our case, the framework differs from standard single clock models. Even though still we have the slow-roll regime, the non-local nature of gravity leads to considerable effects at non-linear level 
		where we have new scale-dependent interactions between different modes depending on the background scalar curvature and the form factor $\Fc_{R}\LF \square\RF $. 
		In the case of local $R^2$ theory interactions between different modes is slow-roll suppressed and leads to the consistency relation \cite{Maldacena:2002vr} but in the case of non-local $R^2$-like inflation, the non-linear evolution of the curvature perturbation gets a significant scale dependence due to non-locality, especially around the scales of Hubble radius exit $k\sim aH$ when $\bar{ R}\gtrsim \Mc_s^2$, leading to violation of consistency relation. This means that we get enhanced scale dependent interactions between the intermediate modes $\Rc$ with $k\sim aH$ and long wavelength modes
		with $k\ll aH$. Being more precise, at the second order action level (\ref{canuac}) we can perform a field redefinition and do canonical normalization that gives the nearly scale invariant power spectrum that matches with the local theory exactly. At the level of the third order action (\ref{3rda}) we invoke the effects of strong scale dependent non-local interactions (especially those interactions with time derivatives of different curvature modes\footnote{{Interactions without any time derivatives of curvature perturbation are slow-roll suppressed, therefore, we have dropped them in our computation since we are interested in only leading order contributions.}}) between different modes in the next to leading order approximation in slow-roll as discussed around (\ref{firstep})-(\ref{examplesl}) and in Appendices~\ref{app.use} and \ref{sec.3point}. It is crucial to emphasize that the key role in the modification of the consistency relation comes from these scale-dependent interaction terms whose effects are negligible in the regime when $\bar{ R}\ll\Mc_{s}^2$ but relevant in the regimes of   $\bar{ R}\gtrsim \Mc_{s}^2$. We can see that the presence of non-local interactions of the different modes in our case gives a result for non-Gaussianity somewhat similar to one in a quasi-single field inflation \cite{Chen:2009zp} where heavy fields with masses of the order of Hubble parameter interact during inflation with curvature perturbation. 
These novel non-local interactions result in various scale dependent shapes  (\ref{A4})-(\ref{A9}) which do not resembles ones known in the literature \cite{Babich:2004gb}.  Indeed, these non-standard interactions yield the violation of the consistency relation.} 
	
{Below we discuss standard shapes of the reduced bi-spectrum which are squeezed, equilateral and orthogonal configurations. As explained above the shapes in question can be discriminated in the non-local $R^2$-like inflation in comparison to the original local $R^2$ model as well as other general scalar field(s) inflationary scenarios \cite{Babich:2004gb,Seery:2005wm,Gao:2008dt} due to non-local nature of gravity.} Considering the general structure of the form-factor in (\ref{formfactors2s}),  we obtain three shapes of reduced bi-spectrum $f_{NL}$ in the leading order approximation as
	\begin{equation}
	\begin{aligned}
	f_{NL}^{sq} & \approx  \Bigg\{\frac{5}{12} \LF 1-n_s \RF -23 \epsilon^2\left[e^{\gamma_S \left(\frac{\bar{ R}}{4 \Mc_s^2}\right)}-1\right] -\frac{4\bar{ R} }{\Mc_s^2}\epsilon^3 e^{\gamma_S \left(\frac{\bar{ R}}{4 \Mc_s^2}\right)} \gamma _S^{(\dagger)}\left(\frac{\bar{ R}}{4 \Mc_s^2}\right)\Bigg\}\Bigg\vert_{k=aH}\,\\ 
	f_{NL}^{eq} & \approx \Bigg\{ \frac{5}{12} \LF 1-n_s \RF -49 \epsilon^2\left[e^{\gamma_S \left(\frac{\bar{ R}}{4 \Mc_S^2}\right)}-1\right]-\frac{9 \bar{ R} }{ \Mc_s^2}\epsilon^3 e^{\gamma_S \left(\frac{\bar{ R}}{4 \Mc_s^2}\right)} \gamma _S^{(\dagger)}\left(\frac{\bar{ R}}{4 \Mc_s^2}\right) \Bigg\} \Bigg\vert_{k=aH}\\ 
	f_{NL}^{ortho} & \approx \Bigg\{ \frac{5}{12} \LF 1-n_s \RF -43 \epsilon^2\left[e^{\gamma_S \left(\frac{\bar{ R}}{4 \Mc_s^2}\right)}-1\right] -\frac{22\bar{ R}}{3 \Mc_s^2}\epsilon^3 e^{\gamma_S \left(\frac{\bar{ R}}{4 \Mc_s^2}\right)} \gamma _S^{(\dagger)}\left(\frac{\bar{ R}}{4 \Mc_s^2}\right) \Bigg\} \Bigg\vert_{k=aH}\,. 
	\end{aligned}
	\label{fnlf}
	\end{equation}
{From the above result for $f_{NL}$ we can deduce that the reduced bispectrum in the non-local $R^2$-like inflation depends not only on the slow-roll parameter $\epsilon$ but also on the background quasi-de Sitter curvature $\bar{ R}$ and the non-local interactions that arise due to the perturbative expansion of form factor $\Fc_{R}\LF \bar{ \square} \RF$ as discussed above in this Section. We can also notice that our results for the  reduced bispectrum in (\ref{fnlf}) indicate a possible running of reduced bispectrum $f_{NL}$. For the time being we postpone the corresponding study for future investigations. }
In (\ref{fnlf}), we have presented only the leading order terms in $\epsilon$ from both the local and non-local contributions (c.f., (\ref{finalfNL})). The parameter space that determines the different shapes of reduced bipsectrum (\ref{fnlf}) is 
	\begin{equation}
		\Biggl\{ \frac{\bar {R}}{4\Mc_{s}^2},\,\gamma_S\LF \frac{\bar{ R}}{4\Mc_{s}^2} \RF,\, \gamma_S^{(\dagger)}\LF \frac{\bar{ R}}{4\Mc_{s}^2} \RF\Biggr\}\Bigg\vert_{k=aH}
	\end{equation}
	From (\ref{fnlf}) we notice that if $\gamma_S\LF \frac{\bar{ R}}{4\Mc_{s}^2} \RF\Big\vert_{k=aH} = \gamma_S^{(\dagger)}\LF \frac{\bar{ R}}{4\Mc_{s}^2} \RF \Bigg\vert_{k=aH}=0 $, we recover the result of local $R^2$ theory in the leading order \cite{Seery:2005wm}.
	
	To see quantitatively the measure of non-Gaussianities, let us consider the following couple of choices of entire function $\gamma_S$ which are compatible with (\ref{gammas}). Moreover we would require an entire function to be compatible with (\ref{condiF}) which reflects an assumption that the higher derivative form-factors do not break the slow-roll approximation scheme (see Appendix~\ref{A2phipsi} for a more detailed explanation).
	
Our first simplest choice of entire function is as below 
	\begin{equation}
\gamma_S\LF \square_s \RF \approx\beta_1 \square_s\LF \square_s-\frac{M^2}{\Mc_{s}^2} \RF\,.
	\label{form1}
	\end{equation}
	The second choice we consider below is higher order in d'Alembertian with a property of $\gamma_S\LF \frac{\bar{ R}}{4\Mc_{s}^2} \RF =0$. This implies the second terms in (\ref{fnlf}) vanish but the 3rd terms can be present and dominate the local contribution depending on the scale of non-locality
	\begin{equation}
	\gamma_S\LF \square_s \RF \approx\beta_2\LF \square_s-\frac{M^2}{\Mc_{s}^2}\RF\LF \square_s-\frac{\bar{ R}_{\ast}}{4\Mc_{s}^2} \RF\LF \square_s+\frac{\bar{ R}}{4\Mc_{s}^2} \RF^6\,.
	\label{form2}
	\end{equation}
	As in the previous Section the approximation sign means that there are higher degree terms in the d'Alembertian operator which would guarantee the supression of the propagator at large momenta but on the other hand provide a freedom choosing signs of arameters $\beta_{1,2}$.

	For the above form-factors (\ref{form1}) and (\ref{form2})  we plot below in figure Fig.~\ref{Staro-plot-fnl} $f_{NL}$ as a function of $\Mc_{s}$ for squeezed, equilateral and orthogonal configurations. Plots are made for $N_\ast=55$ and $\beta_1=1,~\beta_2=-2$. We can read that for $\Mc_s \gtrsim  5.6\times 10^{-5} M_p $, we are well within the bounds of (\ref{lastest.fnl}). If the scale is $5.6\times 10^{-5} M_p < \Mc_{s}< 
	6.5\times 10^{-5} M_p$, we get $f_{NL}\sim \pm O(1)$. We can also notice that all $f_{NL}$ values for different configurations are negative for the case of (\ref{form1}) and positive in the case of (\ref{form2}).
	\begin{figure}[H]
		\centering\includegraphics[height=2.8in]{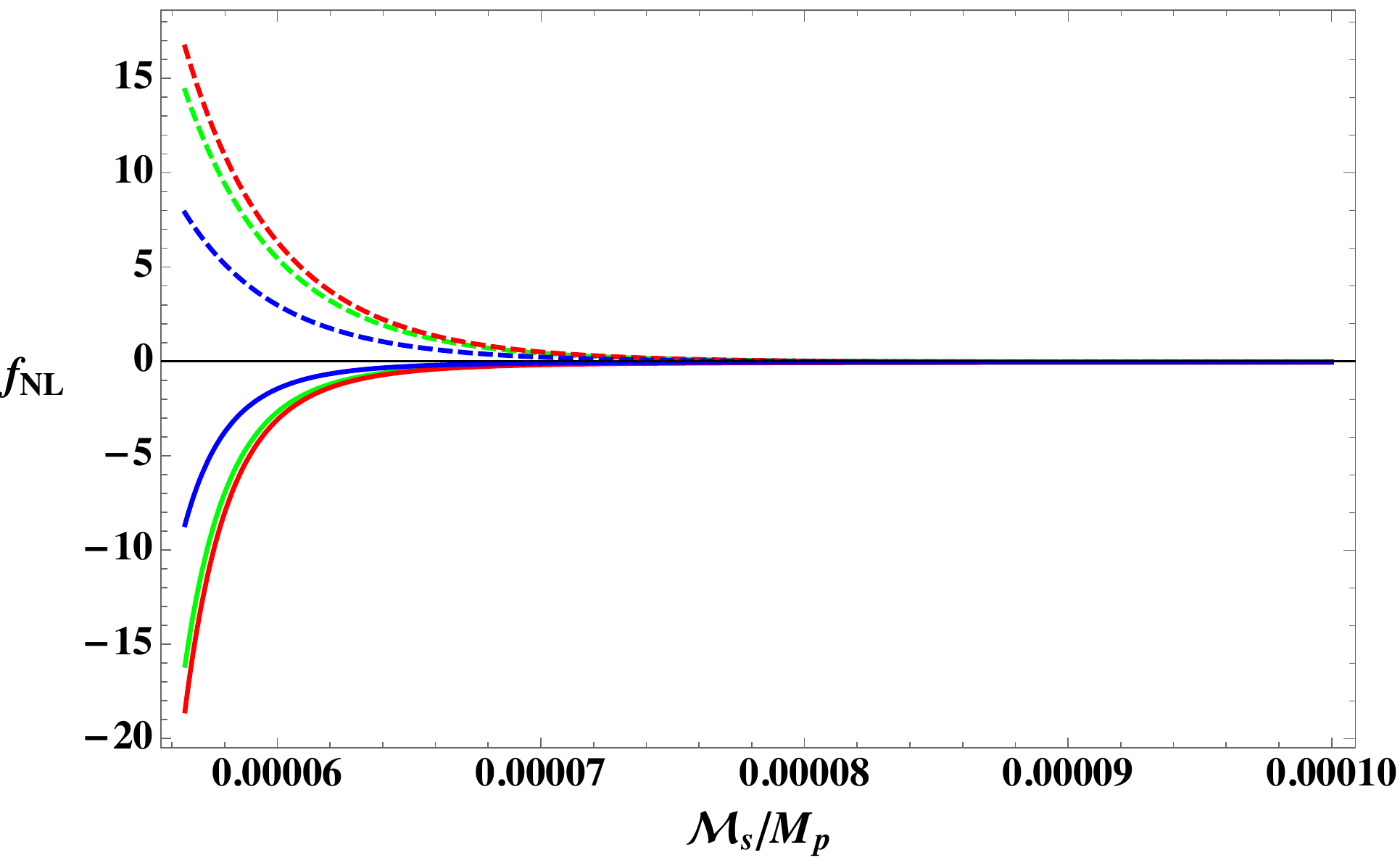}\caption{{In the above plots, $f_{NL}$ versus the scale of non-locality $\Mc_{s}$ (in the units of $M_p$) is depicted for squeezed $k_3\ll k_1=k_2$ (blue), equilateral  $k_1=k_2=k_3$  (red), and orthogonal  $k_1=2k_2=2k_3$ (green) configurations for the entire functions (\ref{form1}) and (\ref{form2}) represented by solid and dashed lines respectively. Here $N_\ast=55$ and $\beta_1=1,~\beta_2=-2$. It can be seen that in the limit $\Mc_{s}\to M_p$, the predictions of the local $R^2$ model are recovered.}}
		\label{Staro-plot-fnl} 
	\end{figure}
	The crucial and significant result here is that the $R^2$-like inflation in AID gravity can give $f_{NL}\sim O(1)$. Moreover, the non-Gaussianity parameter can be of either sign on contrary to the local $R^2$ model which yields only positive values. The large-scale structure observations provide a test-bed for large values of non-gaussianities \cite{Alvarez:2014vva,Liguori:2007sj}. Detecting non-Gaussianities will fix up to some extent the scale of non-locality and the form-factor. 	More importantly, we obtain $f_{NL}^{sq}\sim O(1)$ in our case which has been only known to occur in multifield models of inflation where isocurvature perturbations appear. In our case, although we have only a curvature perturbation here which gets frozen on super-Hubble scales, corrections due to the presence of non-locality can result in $f_{NL}^{sq}\sim O(1)$. Similarly, detectable levels of $f_{NL}^{eq},\, f_{NL}^{ortho}$ have been so far known to be possible in non-canonical models of inflation where the sound speed of curvature perturbations is much less unity. In our case, even though the sound speed of curvature perturbations is unity during inflation, non-local interactions give us $f_{NL}^{eq},\, f_{NL}^{ortho} \sim O(1)$. This makes the $R^2$-like inflation in AID gravity an interesting target with respect to future CMB data.

\section{Conclusions}
	
In this paper we have studied an embedding of the local $R^2$ inflationary solution in the framework of a quadratic in curvatures non-local gravity with analytic infinite derivative form-factors. The essence of the present paper is the computation of the scalar non-Gaussinaities (3-point scalar correlations) and the subsequent attempt to constrain the higher-derivative form-factors based on the existing observational data. 

A strong point of $R^2$-like inflation in the presented non-local gravity setting is that the scalar spectral index $n_s$ remains the same as in the local $R^2$ gravity and thus best fits the current CMB data \cite{Akrami:2018odb}. However, the tensor-to-scalar ratio and the tensor tilt get modified from the standard consistency relation due to the addition of the Weyl tensor squared term with an analytic non-local form-factor in the action \cite{Koshelev:2017tvv}. The analysis reveals that the tensor-to-scalar ratio can acquire any value from the present upper bound all the way down to $10^{-11}$. Furthermore, the tensor tilt is not sign constrained in this model a priori and can take any value in the range $-O(1)<n_t<O(1)$. In particular this disregards possible claims that a sign of the tensor tilt may prove or disprove the inflation as such and moreover a possible blue tensor tilt would highly favor the model outlined in the present paper. In general the performed analysis makes the presented scenario to be a very interesting candidate to be tested within the scope of several future experiments such as BICEP2/Keck, CMB S-4, SO, LiteBIRD and PICO  \cite{Hui:2018cvg,Ade:2018gkx,Abazajian:2016yjj,Abazajian:2019eic,Ade:2018sbj,Hazumi:2019lys,Shandera:2019ufi,Hanany:2019lle}. 

As the main advance in this paper we have computed following the standard methods \cite{Maldacena:2002vr,Weinberg:2005vy} in full scalar non-Gaussianities by means of deriving the 3rd order variation of the action around the $R^2$-like inflationary solution up to the leading order in the slow-roll approximation. We demonstrate that the scalar bi-spectrum contribution comes only from the quadratic in Ricci scalar term.
Assuming suitable form-factors which are compatible with ghost-free conditions of the theory, we have shown that this model is consistent with the current CMB data and leads to predictions that can be tested with future CMB experiments. Namely, our results show that it is possible to obtain $f_{NL}^{sq}\sim O\LF1\RF$ and moreover a sign of the non-Gaussianity parameter is not fixed. Usually any detection of such a large and positive non-Gaussianity is attributed to the presence of multiple scalar fields or to an effect of heavy fields giving rise to non-trivial evolution of curvature perturbations \cite{Chen:2009zp,Byrnes:2014pja}, or to non-slow-roll inflationary scenarios \cite{DeFelice:2013ar}. Our findings present a counterexample to this common point of view since we obtain just a single adiabatic mode of scalar perturbations which can lead to detectable values of non-Gaussianity in the squeezed (local) limit due to the presence of non-local interactions already at the tree level. Furthermore, building a chance to gain large non-Gaussianity parameter of either sign seems to be a unique feature of the present model.

Moreover, $R^2$-like inflation in our setup would also lead to detectable non-Gaussianities in equilateral and orthogonal shapes which is again due to non-local interactions of curvature perturbations when its Fourier modes exit the Hubble radius during inflation.
In turn, this presents an alternative to the occurrence of similar large non-Gaussianities in non-canonical models of inflation \cite{Chen:2006nt} in which the sound speed of curvature perturbations is much less than unity. We emphasize that the sound speed of curvature perturbations in our case is unity during inflation, and different non-Gaussian shapes are solely due to the effect of non-local interactions.

Therefore, any detection of $f_{NL}\sim O\LF1\RF$ or any departure from a standard single field consistency relation can be easily attributed to higher derivative effects in the early Universe. A possible negative values of would be detected non-gaussianity parameters will raise even more the chance for our analytic infinite derivative gravity modification setup to become the favorite one. 
In particular such effects can be put to test in the large-scale structure observations \cite{Alvarez:2014vva,Liguori:2007sj,Meerburg:2019qqi}.
Also our results suggest that the shape of the form-factor $\Fc_{R}\LF \square_s \RF$ can be probed by observations.

In all the instances detectable deviations from so called standard expectations for parameters are subject to the ratio of $\Mc_s/M_p$ which is engaged in the hierarchy with the scalaron mass as expressed in (\ref{scale-hie}) and (\ref{slw-roll})
\[
	M^{2}\ll \Mc_{s}^{2}\lesssim M_{p}^{2}\text{ and }M^2\ll H^2\,.
\label{scale-hie-concl}
\]
For $\Mc_s$ approaching the Planck scale, our results reduce to the ones known from the local $R^2$ model providing no novel detectable signatures of the model. However, the smaller is the scale of the higher derivative gravity modification, the greater is a possible departure towards large non-Gaussianities in particular.
To get a more detailed picture, one needs to constrain the shapes of form-factors which is a well more complicated task requiring more theoretical thinking rather than comparison with observational data. Nevertheless, computing other types of 3-point correlations with the aim to get even more constraints from existing datasets is a very interesting question for forthcoming publications.

	\acknowledgments
	AK is supported by FCT Portugal investigator project IF/01607/2015. 
	This research work was supported by grants  UID/MAT/00212/2019, COST Action CA15117 (CANTATA). KSK is grateful to CMA-UBI for hospitality where part of this research work was done. KSK  and  AM  are  supported  by Netherlands Organization for Scientific Research (NWO) grant number 680-91-119.  AAS is supported by the RSF grant 16-12-10401. We would like to thank G. Calcagni for useful comments. KSK would like to thank J. Marto for numerous enlightening discussions and feedback and S. Maheshwari and P.D. Meerburg for useful discussions. {We thank anonymous referee for useful comments.}
	
	\appendix
	
	\section{Equations of motion, slow-roll approximation and perturbations}
	\label{sec:eom}
	
	The equations of motion for the action (\ref{NC-action}) 
	are \cite{Biswas:2013cha}
	\begin{equation}
	\begin{aligned}E_{\nu}^{\mu}\equiv & -\left[M_{p}^{2}+2\lambda\Fc_R\left(\square_s\right)R\right]G_{\nu}^{\mu}-\frac{\lambda}{2} R\Fc_R\left(\square_s\right)R\delta_{\nu}^{\mu}+2\lambda\left(\nabla^{\mu}\partial_{\nu}-\delta_{\nu}^{\mu}\square\right)\Fc_R\left(\square_s\right)R\\
	& +\lambda\mathcal{K}_{\nu}^{\mu}-\frac{\lambda}{2}\delta_{\nu}^{\mu}\left(\mathcal{K}_{\sigma}^{\sigma}+\tilde{\mathcal{K}}\right)\\
	& +2\lambda(R_{\alpha\beta}+\nabla_\alpha\nabla_\beta)\Fc_W\left(\square_s\right)W_\nu^{\alpha\beta\mu}\\
	& +\frac{\lambda}{2}\delta_{\nu}^{\mu}W^{\alpha\beta\lambda\sigma}{\cal F}_{W}(\square_{s})W_{\alpha\beta\lambda\sigma}-2\lambda W_{\;\alpha\beta\sigma}^{\mu}{\cal {\cal F}}_{W}(\square_{s})W_{\nu}^{\alpha\beta\sigma}\\
	& +\lambda\Omega_{W\nu}^{\mu}-\frac{\lambda}{2}\delta_{\nu}^{\mu}(\Omega_{W\sigma}^{\;\sigma}+\tilde{\Omega}_{W})+4\lambda\Delta_{W\nu}^{\mu}=0\,.
	\end{aligned}
	\label{EoM}
	\end{equation}
	where
	\begin{equation}
	\begin{aligned}\mathcal{K}_{\nu}^{\mu}= &  \frac{1}{\Mc_{s}^{2}}\sum_{n=1}^{\infty}f_{Rn}\sum_{l=0}^{n-1}\partial^{\mu}\square_{s}^{l}R\partial_{\nu}\square_s^{n-l-1}R\,,\quad
	\tilde{\mathcal{K}}=   \sum_{n=1}^{\infty}f_{Rn}\sum_{l=0}^{n-1}\square_{s}^{l}R\square_{s}^{n-l}R\,\\
	\Omega_{W\nu}^{\mu}= &\frac{1}{\Mc_{s}^{2}}  \sum_{n=1}^{\infty}f_{W{n}}\sum_{l=0}^{n-1}\square_s^lW_{\:\:\beta\lambda\sigma;\nu}^{\alpha}\square_s^{n-l-1}W_{\alpha}^{\;\beta\lambda\sigma;\mu},~
	\tilde{\Omega}_{W}=\sum_{n=1}^{\infty}f_{W{n}}\sum_{l=0}^{n-1}\square_s^lW_{\:\:\beta\lambda\sigma}^{\alpha}\square_s^{n-l}W_{\alpha}^{\;\beta\lambda\sigma}\\
\Delta_{W\nu}^{\mu}= & \frac{1}{\Mc_{s}^{2}}  \sum_{n=1}^{\infty}f_{W{n}}\sum_{l=0}^{n-1}[\square_s^lW_{\quad\sigma\alpha}^{\lambda\beta}\square_s^{n-l-1}W_{\lambda;\nu}^{\;\mu\sigma\alpha}-\square_s^lW_{\quad\sigma\alpha;\nu}^{\lambda\beta\;\;}\square_s^{n-l-1}W_{\lambda}^{\:\mu\sigma\alpha}]_{;\beta}\,.
	\end{aligned}
	\label{KmnRFR-1}
	\end{equation}
	In the above expressions $\nabla$ and $;$ denote covariant derivatives.
	
\subsection{Slow-roll approximation}

During the quasi-de Sitter expansion, $\bar{R}$ is nearly constant that means
\begin{eqnarray}
H & \approx & \const\,,\quad a\approx a_{0}e^{Ht}\,,\\
R_{\mu\nu\rho}^{\sigma} & \approx & \frac{R}{12}(\delta_{\nu}^{\sigma}g_{\mu\rho}-\delta_{\rho}^{\sigma}g_{\mu\nu})\,,\quad R_{\nu}^{\mu}\approx\frac{R}{4}\delta_{\nu}^{\mu},\quad R\approx\const.\label{nearlyds}
\end{eqnarray}
In terms of the conformal time, the de Sitter space-time is defined by
\begin{equation}
a\approx-\frac{1}{H\tau}\,,\quad\Hc\approx-\frac{1}{\tau}\:\text{ and }\:\Hc'\approx\Hc^{2}\approx\frac{1}{\tau^{2}}\,.\label{nearlydstau}
\end{equation}
Quasi-de Sitter space is a slight departure from the exact de Sitter which can be defined by the slow-roll parameter 
\begin{equation}
\epsilon = -\frac{\dot{H}}{H^2}\approx \frac{M^2}{6H^2}=\frac{1}{2N}\ll 1\,. 
\end{equation}
where $H=\frac{\dot{a}}{a}$ is the Hubble parameter, $M$ is the scalaron mass and $N=\ln a_f-\ln a$ is the number of e-folds during inflation counted from its end 
($a=a_f$) backwards in time. The slow-roll parameter above is for the local $R^2$ inflation (\ref{scale-fac}). 
The following relations are useful in many of our computations
\begin{equation}
\epsilon=-\frac{\dot{H}}{H^{2}}=1-\frac{\mathcal{H}^{\prime}}{\mathcal{H}^{2}}\,,\quad\epsilon^{\prime}=2\mathcal{H}\left(1-\epsilon\right)^{2}-\frac{\mathcal{H}^{\prime\prime}}{\mathcal{H}^{2}}\,,\label{epsilon}
\end{equation}
where $H=\frac{\dot{a}}{a},\,\mathcal{H}=\frac{a^{\prime}}{a}$ are
the Hubble parameter in the cosmic and conformal time respectively

\begin{equation}
\begin{aligned}\bar{R}= & \frac{6\mathcal{H}^{2}}{a^{2}}\left(\mathcal{H}^{\prime}+\mathcal{H}^{2}\right)\sim\frac{12\mathcal{H}^{2}}{a^{2}}\,,\\
\bar{R}^{\prime}= & \frac{6\mathcal{H}^{2}}{a^{2}}\left(\frac{\mathcal{H}^{\prime\prime}}{\mathcal{H}^{2}}-2\mathcal{H}\right)= \frac{6\mathcal{H}^{2}}{a^{2}}\left(-4\mathcal{H}\epsilon-\epsilon^{\prime}\right)
\approx  -8\bar{R}\mathcal{H}\epsilon\approx-4M^2\mathcal{H}\,.
\end{aligned}
\label{Rprime}
\end{equation}
We note that slow-roll would also mean $M^2=\frac{6\mathcal{H}^{2}}{a^{2}}\epsilon=\frac{1}{2}\bar{R}\epsilon\to0$
which can be easily deduced using (\ref{Rprime}) and (\ref{ansatz-1})
in the limit $\epsilon\ll1$. Since $\bar{R}^{\prime}\sim{O}\left(\epsilon\right)$,
it implies $\bar{R}^{\prime\prime}\sim{O}\left(\epsilon^{2}\right)\approx0$.

\subsection{On $\Phi+\Psi\approx 0$ during quasi-de Sitter expansion}
\label{A2phipsi}
One of the prime result of \cite{Koshelev:2017tvv} is that second order action of scalar perturbations in AID gravity remains the same as the one in the  local $R^2$ gravity, around the inflationary solution (\ref{scale-fac}) that satisfies (\ref{ansatz-1}), up to the leading order in slow-roll approximation. Here we discuss this point briefly and heuristically show that non-local corrections at the second order level are indeed negligible. The exact from of second order action of AID gravity around FLRW space times that satisfy the background equation (\ref{ansatz-1}) reads as \cite{Koshelev:2017tvv} 
\begin{equation}
\delta^2S=\intdg{4}\left[\delta_{local}+\frac{1}{2}\zeta \Zc_2(\bar\square_s)\zeta+\frac{1}{2}\delta W_{\mu\nu\rho\sigma}\Fc_W(\bar\square_s)\delta W^{\mu\nu\rho\sigma}\right]\,, 
\label{delta2Sinf}
\end{equation}
where $\delta_{local}$ is the second order variation of an action
\begin{equation}
S_{local}=\intdg{4}\left(\frac {M_P^2}2R+\frac1 2\Fc_1R^2\right)\,,
\label{actionour4inflocal}
\end{equation}
and 
\begin{equation}
\zeta = \frac{1}{\Mc_{s}^2}\LT\LF \bar{ \square}-M^2 \RF\delta R+ \delta{ \square}\bar{R} \RT\,.
\end{equation}
Moreover it was shown by solving perturbed trace equation of AID gravity that $\zeta=0$ in the leading order in the slow-roll approximation, and this is exactly equivalent to $\Phi+\Psi\approx 0$. Since the first order variations of the Weyl tensor are proportional to $\Phi+\Psi\approx 0$, there will be no scalar contributions from the Weyl tensor term. As a result, we get the scalar power spectra like in the local $R^2$ model up to the leading order corrections as we discussed in Section~{\ref{sec2a}} and we can also easily get it from (\ref{delta2Sinf}).

Now, we can heuristically translate this result into a condition on the form-factor $\Fc_R\LF \bar{ \square}_s \RF$ by computing the contributions of the second term in (\ref{delta2Sinf})
in the next to leading order in the slow-roll approximation and showing them to be negligible compared to the contribution of the local term (\ref{actionour4inflocal}). To show this, we consider $\Phi+\Psi\approx 0$ and $\bar{ \square}\Psi \approx M^2\Psi$. Now calculating slow-roll corrections to the second term in (\ref{delta2Sinf}), we obtain
\begin{equation}
\begin{aligned}
\int d^4x \sqrt{-g}\zeta\Zc_{2}\LF \bar{ \square}_s \RF\zeta \approx & \frac{M^4}{\Mc_{s}^4} \intdg4 \delta R\Zc_{2}\LF \bar{ \square}_s \RF \delta R \\ 
\approx & \frac{M^4}{\Mc_{s}^4}  \intdg4 \delta R\Fc_{R}^{(\ddagger)}\LF \frac{M^2}{\Mc_{s}^2} \RF \delta R\,, 
\end{aligned}
\label{zcorr}
\end{equation}
where we applied the approximation $\bar{ R}\approx\const$, and $\Fc_{R}^{(\ddagger)}\LF \frac{M^2}{\Mc_{s}^2} \RF$ is the second derivative of the form-factor evaluated at the value of scalaron mass square. To declare that the contribution (\ref{zcorr}) is negligible compared to the local term from the quadratic part in the local action (\ref{actionour4inflocal}), we require 
\begin{equation}
\Fc_{1}=\frac{M_p^2}{6M^2} \gg \frac{M^4}{\Mc_{s}^4} \Fc_{R}^{(\ddagger)}\LF \frac{M^2}{\Mc_{s}^2} \RF\,.
\label{condiF}
\end{equation}
Considering the hierarchy of scales as (\ref{scale-hie}), we can easily satisfy the above condition if the second derivative of the form-factor evaluated at $M^2$ satisfy $\Fc_{R}^{(\ddagger)}\LF \frac{M^2}{\Mc_{s}^2} \RF \ll \frac{M_p^2\Mc_{s}^4}{6M^6}$.

\section{Useful relations for the computation of bi-spectrum}
\label{app.use}
	
	In this section, we provide computations of perturbed curvature quantities and action of infinite derivatives on them within the slow-roll approximations. Many of the following simplifications were used in computing the 3-point correlations in the previous section.
	
	Perturbation functions are not space-homogeneous. It is convenient
	to perform a spatial Fourier transform which for some function $\varphi(\tau,\vec{x})$
	in the flat FLRW metric reads
	\[
	\varphi(\tau,\vec{x})=\int\varphi(\tau,\vec{k})e^{i\vec{k}\vec{x}}d\vec{k}\,.
	\]
	Using the above Fourier representation, we can write
	\[
	\bar{\square}\varphi(\tau,\vec{x})=-\int\frac{1}{a^{2}}(\pd_{\tau}^{2}+2\Hc\pd_{\tau}+k^{2})\varphi(\tau,\vec{k})d\vec{k}\,,
	\]
	The first and second variations of the d'Alembertian operator are
	\begin{equation}
	\begin{aligned}\delta\square= & 2\Psi\bar{ \square}+2\dot{\Psi}\pd_{t}-\frac{2}{a^2}\nabla\Psi\cdot\nabla\,\\
	\delta^{2}\square= & 4\Psi^2\bar{ \square}-\frac{8}{a^2}\Psi\nabla\Psi.\nabla + 8\Psi\dot{ \Psi}\pd_t\,.
	\end{aligned}
	\label{VaryBox}
	\end{equation}
	Below we provide the useful list of perturbations of scalar curvature up to the 3rd order 
	\begin{equation}
	\begin{aligned}
	\delta R &= 2\LF \bar{R}+3\bar{\square}_k \RF \Psi \,\\
	\delta^{(2)}R &=  \frac{6}{a^{2}}\Bigg[ \nabla \Psi\cdot \nabla\Psi  -\Psi^{\prime^{2}} +4a^2\Psi\square\Psi +\frac{2}{3}a^2\Psi^2\bar{R} \Bigg]\,\\
	\delta^{(3)} R &  = \frac{12\Psi}{a^2} \Bigg[3 \nabla \Psi\cdot \nabla \Psi-\ 3\Psi^{\prime 2}+ 6a^2\Psi\square\Psi +\frac{2}{3}a^2\Psi^2\bar{R} \Bigg]    \\ 
	\delta\LF \sqrt{-g}R \RF &=  a^4 \Bigg[ \delta R-4\Psi\bar{R} \Bigg] \\
	\delta^{(2)}\LF \sqrt{-g}R\RF  &= 6a^2\Bigg[\nabla\Psi\cdot\nabla\Psi-\Psi^{\prime {2}}\Bigg]\\
	\delta^{(3)}\left(\sqrt{-g}R\right) &= 12 a^2\Psi  \Bigg [\nabla \Psi\cdot \nabla\Psi  -\Psi^{\prime 2} \Bigg]\,.
	\end{aligned}
	\label{Two-3Rvary}
	\end{equation}
	Below we derive some recurrent relations when d'Alembertian operators act on perturbed quantities 
	\begin{equation}
	\begin{aligned}
	\square_s \delta\square_s \bar{R} & \approx \bar{\square}_s\LF \frac{2M^2}{\Mc_s^2}\bar{R}\Psi- \frac{2\bar{R}}{\Mc_{s}^2}H\epsilon\dot{\Psi} \RF  \\ 
	&	\approx  \LF\frac{2M^2 }{\Mc_s^2}\RF^2\bar{R}\Psi  -\frac{2\bar{R}}{\Mc_{s}^2}H\epsilon \LF \frac{M^2}{\Mc_{s}^2}+\frac{\bar{R}}{4\Mc_{s}^2} \RF \dot{\Psi}\\ 
	\implies \bar{\square}_s^n\delta\square_s \bar{R} &  \approx \LF \frac{2M^2}{\Mc_s^2} \RF^{(n+1)} \bar{R}\Psi-\frac{2\bar{R}}{\Mc_{s}^2}H\epsilon \LF \frac{M^2}{\Mc_s^2}+\frac{\bar{R}}{4\Mc_{s}^2}\RF^{n} \dot{\Psi} \\ 
	\implies \Oc\LF \square_s \RF\delta\square_s \bar{R} & \approx \frac{2M^2}{\Mc_{s}^2} \Oc\LF \frac{2M^2}{\Mc_{s}^2} \RF\bar{R}\Psi - \frac{2\bar{R}H\epsilon}{\Mc_{s}^2} \Oc\LF \frac{M^2}{\Mc_s^2}+\frac{\bar{R}}{4\Mc_{s}^2} \RF\dot{ \Psi}\,.
	\end{aligned}
	\end{equation}
	In the same way we can compute 
	\begin{equation}
	\begin{aligned}
	\bar{\square}_s\delta R & \approx 2\bar{\square}_s\LF\bar{R} \Psi \RF  \\ 
	& \approx 2\frac{2M^2}{\Mc_s^2}\bar{R}\Psi +2\frac{2\bar{R}}{\Mc_{s}^2}H\epsilon \dot{\Psi} \\ 
	\implies \bar{\square}_s^n \delta R & \approx 2\LF \frac{2M^2}{\Mc_{s}^2} \RF^n\bar{R}\Psi+  \frac{2\bar{R}}{M^2}H\epsilon \LF\frac{2M^2}{\Mc_{s}^2}\RF^n\dot{ \Psi} + \frac{4\bar{R}H\epsilon}{\Mc_{s}^2\LF \frac{M^2}{\Mc_s^2}+\frac{\bar{R}}{4\Mc_{s}^2} \RF} \LF \frac{M^2}{\Mc_s^2}+\frac{\bar{R}}{4\Mc_{s}^2} \RF^n \dot{\Psi} \,. 
	\end{aligned}
	\label{listzfz2-1}
	\end{equation}
	Using (\ref{listzfz2-1}) we can deduce the following
	\begin{equation}
	\begin{aligned}
	\Oc\LF \bar{\square}_s \RF \delta R & \approx \Oc\LF \frac{2M^2}{\Mc_s^2} \RF  2\bar{R}\Psi + \frac{2\bar{R}}{M^2}H\epsilon \LT \Oc\LF\frac{2M^2}{\Mc_{s}^2}\RF-\Oc(0)\RT \dot{\Psi} \\& +16H\epsilon\LT \Oc\LF \frac{M^2}{\Mc_s^2}+\frac{\bar{R}}{4\Mc_{s}^2} \RF -\Oc(0)\RT \dot{\Psi}\,.
	\end{aligned}
	\label{listzfz2}
	\end{equation}
	Based on generic formula derived
	in \cite{Koshelev:2017tvv}
	\begin{equation}
\begin{aligned}
\int d^{4}x\sqrt{-\bar{g}}\bar{ R}\delta\square_s X &=\int d^{4}x\sqrt{-\bar{g}}\LT\delta\square_s\bar{R}-\frac{1}{2\Mc_s^2}\bar{R}h\LF\bar{\square}-M^2\RF\RT X\,. 
\end{aligned}
\label{term3de-1}
\end{equation}
	for any $X$, we derive the following generic result which we often used in our computation of the 3rd order action.
	Computing explicitly the first term we get 
	\begin{equation}
	\begin{aligned}
	\int d^4x\sqrt{-\bar{g}}\delta\square_{s}\bar{R} X = & \frac{1}{\Mc_{s}^2}\int d^4x \sqrt{-\bar{g}} \LT  2M^2\Psi\bar{R}+2\dot{\Psi}\dot{\bar{R}} \RT  X\\ 
	\approx &  \frac{1}{\Mc_{s}^2}\int d^4x \sqrt{-\bar{g}} \LT  M^2\delta R+2\dot{\bar{R}}\dot{\Psi} \RT  X\,.
	\end{aligned}
	\end{equation}
	Substituting the trace of $h= \bar{g}^{\mu\nu}h_{\mu\nu}= 2\Phi-6\Psi \approx -8\Psi $ in (\ref{term3de-1}), we arrive at
	\begin{equation}
	\begin{aligned}
	&\int d^{4}x\sqrt{-\bar{g}}\bar{ R}\delta\square_s X  \approx \int d^4x\sqrt{-\bar{g}} \frac{1}{\Mc_s^2} \LT M^2\delta R+2\dot{\bar{R}}\dot{ \Psi}+2 \LF \bar{ \square}-M^2 \RF \delta R  \RT X\, \\ 
	  \approx& -6\int d^4x\sqrt{-\bar{g}} \frac{1}{\Mc_s^2} \LT \dot{\bar{R}}\dot{ \Psi}-M^2\bar{R}\Psi\RT X 
	  \approx \frac{6}{\Mc_{s}^2}\int d^4x\sqrt{-\bar{g}} \LT 2\bar{ R}H\epsilon\dot{ \Psi} +M^2\bar{ R}\Psi\RT X\,.
	\end{aligned}
	\label{z1delR}
	\end{equation}
	where we have used $\bar{ \square}\delta R= 2M^2\delta R - 4\dot{\bar{R}}\dot{ \Psi}$.

\section{Computation of 3-point correlations}
\label{sec.3point}

In this section, we present detailed computations of the 3-point correlation (\ref{3-point-f}). For this, we first need to expand the action (\ref{NC-action}) up to the cubic order in curvature perturbation ($\Rc$) which would be given by 
\begin{equation*}
\begin{aligned} S & = \delta^{(2)}S_{(S)}+\delta^{(3)} S_{(S)} \,, 
\end{aligned}
\end{equation*}
where 
\begin{equation}
\begin{aligned} 
	\delta^{(3)}S_{(S)} & = \frac{M_p^2}{2}\int d^4x\,\delta^{(3)}\LF \sqrt{-g}R \RF+ 
\frac{1}{2}\int d^{4}x\,\Bigg[ \delta^{3}\left(\sqrt{-g}R\right)\Fc_1\bar{R}
+\sqrt{-\bar{g}}\Fc_1\bar{R} \delta^{(3)}R\\& 
+\Fc_{1}\delta^{(2)}\LF \sqrt{-g}R \RF\delta R  +\Fc_{1}\delta\LF\sqrt{-g}R\RF  \delta^{(2)} R\\& 
+ \delta^{(2)}\LF \sqrt{-g}R \RF \Bigg(\Fc_{R}\left(\square_{s}\right)-\Fc_{1}\Bigg)\delta R  {+\delta\LF\sqrt{-g}R\RF  \Bigg(\Fc_{R}\left(\square_{s}\right)-\Fc_{1}\Bigg) \delta^{(2)} R}\\ & +
\delta^{(2)}\LF\sqrt{-g}R\RF\delta\mathcal{F}_{R}\left(\square_{s}\right)\bar{R}
+\sqrt{-\bar{g}}\bar{R}\delta\Fc_{R}\left(\square_{s}\right)\delta^{(2)}R
+\delta\LF\sqrt{-g} R\RF\delta\Fc_{R}\left(\square_{s}\right)\delta R  \\&
+ \delta\LF\sqrt{-g} R\RF\delta^{\left(2\right)}\Fc_{R}\left(\square_{s}\right)\bar{R}
+\sqrt{-\bar{g}}\bar{R}\delta^{\left(2\right)}\Fc_{R}\left(\square_{s}\right)\delta R+\sqrt{-\bar{g}}\bar{R}\delta^{(3)}\Fc_R\LF\square_{s}\RF\bar{R}\\
& +\sqrt{-\bar{g}}\delta^{(2)}W_{\mu\nu\alpha\beta}\Fc_{W}\left(\square_{s}\right)\delta W^{\mu\nu\alpha\beta}\\ &+\sqrt{-\bar{g}}\delta W_{\mu\nu\alpha\beta}\mathcal{F}_{W}\left(\square_{s}\right)\delta^{(2)}W^{\mu\nu\alpha\beta}+\sqrt{-\bar{g}}\delta W_{\mu\nu\alpha\beta}\delta\Fc_{W}\left(\square_{s}\right)\delta W^{\mu\nu\alpha\beta}\Bigg]\,.
\end{aligned}
\label{eq:3rdac}
\end{equation}
Here we performed integration by parts and used the background solution (\ref{ansatz-1}). The contribution from the Einstein-Hilbert term is negligible because the $R^2$ term dominates during inflation. 

Since $\Phi+\Psi\approx 0$ during inflation, we compute (\ref{eq:3rdac}) up to the cubic order in $\Psi \approx \epsilon \Rc$ in the leading order slow-roll approximation. 
From \cite{Koshelev:2016xqb}, we know that the first variation of
Weyl term only contributes to tensor perturbations and does not lead to scalar perturbations because $\Phi+\Psi \approx 0$ during inflation. Notice that all the terms in (\ref{eq:3rdac}) contains at least one first order variation of the Weyl tensor. Therefore, these contributions can be approximated by zero for the scalar 3-point correlation. 
As a result, the cubic order scalar contributions can only arise from the quadratic Ricci scalar term in the action. 

Moreover, for the computation of 3-point correlations (\ref{3-point-f}) we use the mode functions evaluated from the second order action of curvature perturbations evaluated is the de Sitter approximation (\ref{canuac}) which we recall here again expressing in terms of curvature perturbation $\Rc$ as 
\begin{equation}
	\delta^{(2)}S_{(S)} = \frac{M_p^2}{2}\epsilon\int d\tau d^3k a^4 \Rc\LF \bar{\square}-M^2 \RF \Rc \,.
    \label{Rceq}
\end{equation}
The above action in the quasi-de Sitter approximation gives two-point correlations possessing a nearly scale invariant power spectrum which is Gaussian. Non-Gaussianities arise from 3-point correlations which are computed from the 3rd order action in the next to leading order in the slow-roll approximation.   
In the computation of the 3-order action it is important to eliminate terms  proportional to the equation of motion for $\Rc$ (it can be obtained by varying the action (\ref{Rceq})) via a suitable field redefinition of $\Rc$. This is step is crucial to extract the leading non-linear contributions to curvature perturbations \cite{Maldacena:2002vr,Seery:2005wm}.  In the local $R^{2}$ gravity, it is known that non-Gaussianities are small \cite{Maldacena:2002vr}. However, in the non-local
gravity we have non-local contributions to the bi-spectrum which can be read from (\ref{eq:3rdac}). Note that in the calculation of bi-spectrum, we can perform the computation using the leading order in the slow-roll approximation. 

In (\ref{eq:3rdac}) there are several terms involving variations of the form-factor which can be expanded as 
\begin{equation}
	\begin{aligned}\delta\mathcal{F}_{R}\left(\square_{s}\right)= & \sum_{n=1}^{\infty}f_{n}\sum_{a+b=n-1}\square_{s}^{a}\delta\square_{s}\square_{s}^{b}\,,\\
		\delta^{(2)}\mathcal{F}_R\left(\square_{s}\right)= & \sum_{n=1}^{\infty}f_{n}\sum_{a+b=n-1}\square_{s}^{a}\delta^{(2)}\square_{s}\square_{s}^{b}+\sum_{n=2}^{\infty}f_{n}\sum_{a+b+c=n-2}\square_{s}^{a}\delta\square_{s}\square_{s}^{b}\delta\square_{s}\square_{s}^{c}\,,\\
\delta^{(3)}\Fc_R(\square_{s})= & \sum_{n=1}^{\infty}f_{n}\sum_{a+b=n-1}\square_{s}^{a}\delta^{(3)}\square_s\square_{s}^{b}+\sum_{n=2}^{\infty}f_{n}\sum_{a+b+c=n-2}\square_{s}^{a}\delta^{(2)}\square_{s}\square_{s}^{b}\delta\square_{s}\square_{s}^{c}\\
& +\sum_{n=2}^{\infty}f_{n}\sum_{a+b+c=n-2}\square_{s}^{a}\delta\square_{s}\square_{s}^{b}\delta^{(2)}\square_{s}\square_{s}^{c}
\\&+\sum_{n=3}^{\infty}f_{n}\sum_{a+b+c+d=n-3}\square_{s}^{a}\delta\square_{s}\square_{s}^{b}\delta\square_{s}\square_{s}^{c}\delta\square_{s}\square_{s}^{d}\,.
\end{aligned}
\label{FRvariations}
\end{equation}
Below, we perform the term by term computation of the 3rd order action (\ref{eq:3rdac})  in terms of the curvature perturbation $\Rc$ and the evaluation of 3-point correlation functions\linebreak $\langle\Rc\left(\mathbf{k_{1}}\right)\Rc\left(\mathbf{k_{2}}\right)\Rc\left(\mathbf{k_{3}}\right)\rangle$ in each case. Note that in all the following calculations, we only write the interaction terms which are leading order in slow-roll. Also in the calculations we encounter terms proportional to the quantity 
\begin{equation}
\Fc_{R}\LF \frac{2M^2}{\Mc_{s}^2}\RF-\Fc_{1} \approx 0 \,, 
\end{equation}
which is found to be negligible (for the form-factors of the form (\ref{formfactors2s})) compared to remaining terms which contribute significantly to the bi-spectrum. Therefore, we drop all these terms for economic reasons. Also we do not write the interaction terms which only lead to imaginary values of 3-point correlations, as by definition the 3-point correlation function is a real quantity 
(\ref{3-point-f}). 


First we compute the local contribution of the 3rd order action  (\ref{eq:3rdac}) which is the 3rd variation of the local action below 
	\begin{equation}
	    S_{\textrm{local}} = \int d^4x \sqrt{-g}\LF M_p^2R+\Fc_1R^2 \RF\,. 
	\end{equation}
	\begin{equation}
	\begin{aligned}
	\delta^{(3)}S_{\textrm{local}}= & \,\, \int d^4x \Bigg\{\Bigg[\delta^{(3)}\LF \sqrt{-g} R \RF  +\sqrt{-\bar{g}}\delta^{(3)}R\Bigg]\Fc_1 \bar{R} \\ &+ \Bigg[\delta^{(2)}\LF \sqrt{-g}R \RF\delta R+ \delta\LF\sqrt{-g}R\RF \delta^{(2)} R\Bigg]\Fc_1\Bigg  \} \\ 
	 \approx \,\, &-2 M_p^2\LF\epsilon^2 +\frac{3}{4}\epsilon^3 \RF \int  d\tau d^3x   a^2 \Bigg[8\Rc \nabla\Rc\cdot\nabla\Rc - 8\Rc\Rc^{\prime 2}\Bigg] \\ & 
	 -4M_p^2\epsilon^3\int d\tau d^3x a^4  \bar{ R}\Rc^3- 8\epsilon \int d\tau d^3x a^4 \Rc^2\frac{\partial\Lc_2}{\partial\Rc}\,+ O\LF \epsilon^3 \RF, 
		\end{aligned}
	\label{c1f}
	\end{equation}
where $\frac{\pd\Lc_2}{\pd\Rc}$ is the term proportional to the equation of motion of $\Rc$ following from (\ref{Rceq}). This term can be eliminated by a field redefinition of $\Rc\to \Rc + 4\epsilon\Rc^2 $ which leads to a modification of the Gaussian term as 
\begin{equation}
    \delta^{(2)}S \to \delta^{(2)}S+4\epsilon\int d\tau d^3x a^4 \Rc^2\frac{\pd\Lc_2}{\pd\Rc}\,. 
    \end{equation}
This term leads to local contributions to the bi-spectrum which are (\ref{A1}) to (\ref{A3})
\begin{equation}
T_1 \supseteq -2\epsilon-\frac{3\epsilon^2}{4}\,,\qquad  T_2 \supseteq 2\epsilon+\frac{3\epsilon^2}{4}\,, \qquad T_3 \supseteq -\frac{\epsilon^2}{2}\,. 
\end{equation}

Now we calculate 3-point correlations from the following non-local term which does not involve variation of the form-factor 
\begin{equation}
\begin{aligned}
\int d^4x  \Bigg\{\delta^{(2)}\LF \sqrt{-g}R \RF & \Bigg[ \Fc_R\left(\square_{s}\right) -\Fc_{1} \Bigg]\delta R+ \delta\LF\sqrt{-g}R\RF \Bigg[ \Fc_R\left(\square_{s}\right) -\Fc_{1} \Bigg] \delta^{(2)} R\Bigg\}
\\ \approx &\,\, 64\epsilon^4 T_{\textrm{NL}}\int d\tau d^3x a^2 \bar{ R}\Hc\Rc^\prime\Rc^2 \,\\
\end{aligned}
\label{c2f}
\end{equation}
where 
\begin{equation*}
\begin{aligned}
     T_{\textrm{NL}}&= \LT \Fc_{R}\LF \frac{M^2}{\Mc_s^2}+\frac{\bar{R}}{4\Mc_{s}^2} \RF -\Fc_{1}\RT \,. 
    \end{aligned}
\end{equation*}
This term contributes to (\ref{A4}) 
\begin{equation}
T_4 \supseteq \frac{8\bar{ R}}{M_p^2}\epsilon^3 T_{\textrm{NL}}\,. 
\end{equation}
 
Let us consider the following term of  (\ref{eq:3rdac}) which involve first variation of the form-factor $\Fc_{R}\LF \square_s \RF$ 
	\begin{equation}
	\begin{aligned}\int d^{4}x & \Bigg[\delta^{(2)}  \LF\sqrt{-g}R\RF\delta\mathcal{F}_{R}\left(\square_{s}\right)\bar{R}
	+ \sqrt{-\bar{g}}\bar{R}\delta\Fc_{R}\left(\square_{s}\right)\delta^{(2)}R \Bigg]\\  = & \int d^{4}x \Bigg[\delta^{(2)}  \LF\sqrt{-g}R\RF\Zc_{1}\LF \square_s \RF \delta\square_{s}  \bar{ R} + \sqrt{-\bar{g}}\bar{ R}\delta\square_s\Zc_{1}\LF \square_s \RF\delta^{(2)}R\Bigg] \\
	= & -24\epsilon^4 T_{\textrm{NL}}\int d^{4}x  8\Hc\Rc^\prime\Bigg[  \nabla\Rc\cdot\nabla\Rc-\Rc^{\prime 2} + a^2\bar{ R}\Rc^2 \Bigg]\,.
	\end{aligned}
	\label{c3f}
	\end{equation}
	Passing from the first line to the second one, we have carried integration by parts and used the background solution $\bar{ \square}\bar{R}= M^2\bar{R}$. Passing from the second line to the third one, we have substituted expressions from (\ref{Two-3Rvary}) and the result derived in (\ref{z1delR}). This term gives the following contribution to the bi-spectrum (\ref{A4})-(\ref{A6}) 
	\begin{equation}
	T_4\supseteq -\frac{24\bar{ R}}{M_p^2}\epsilon^3 T_{\textrm{NL}}\,,\quad T_5\subseteq -\frac{2\bar{ R}}{M_p^2}\epsilon^3 T_{\textrm{NL}}\,,\quad  T_6 \supseteq \frac{2\bar{ R}}{M_p^2}\epsilon^3 T_{\textrm{NL}}\,. 
	\end{equation}

Let us now consider the following non-local contribution of (\ref{eq:3rdac}) 
\begin{equation}
\begin{aligned}
&\int d^4x \delta\LF \sqrt{-g} R\RF \delta \Fc_R\LF \bar{ \square}_s \RF\delta R  \approx  -\int d^4x\sqrt{-\bar{g}} \delta R  \delta\square_s\Zc_{1}\LF \bar{ \square}_s \RF\delta R\,\\
\approx &  \,\,128\epsilon^4T_{\textrm{NL}}\int d\tau d^3x a^2 \Rc \Hc\Bigg[\frac{\bar{R}}{2}\Rc\Rc^\prime+  \frac{2}{a^2}\Rc^\prime\Rc^{\prime\prime}-\frac{2}{a^2}\nabla\Rc\cdot\nabla\Rc^\prime\Bigg]\,. 
\end{aligned}
\label{c4f}
\end{equation}
The contributions form this term are (\ref{A4}), (\ref{A8}) and (\ref{A9}) 
\begin{equation}
T_4\supseteq \frac{8\bar{ R}}{M_p^2}\epsilon^3T_{\textrm{NL}},\, \quad T_8\supseteq \frac{8\bar{ R}}{3M_p^2}\epsilon^3T_{\textrm{NL}}\,,\quad  T_9 = -\frac{8\bar{ R}}{3M_p^2}\epsilon^3T_{\textrm{NL}} \,. 
\end{equation}

Let us now consider the following two non-local terms which require second variation of the form-factor $F_R\LF \square_s \RF$. They can be written as two kinds of variations in the following way
\begin{equation}
\begin{aligned}
\int d^{4}x\sqrt{-\bar{g}}\bar{R}\delta^{(2)}\Fc_{R}\left(\bar{\square}_{s}\right)\delta R= &\int d^{4}x\sqrt{-\bar{g}}\bar{R}\left[\delta^{(2)}\Fc_{R}\left(\bar{\square}_{s}\right)\Bigg\vert_{\delta^{(2)}\square_{s}}\right]\delta R\\ & +\int d^{4}x\sqrt{-\bar{g}}\bar{R}\left[\delta^{(2)}\Fc_{R}\left(\bar{\square}_{s}\right)\Bigg\vert_{\delta\square_{s}\delta\square_{s}}\right]\delta R\,. 
\end{aligned}
\label{c5f}
\end{equation}
Calculating the first part in (\ref{c5f}), we obtain
\[
\begin{aligned}
	&\int d^{4}x\sqrt{-\bar{g}}\bar{R}\left[\delta^{(2)}\Fc_{R}\left(\bar{\square}_{s}\right)\Bigg\vert_{\delta^{(2)}\square_{s}}\right]\delta R\\
	= & \int d^{4}x\sqrt{-\bar{g}}\bar{R}\sum_{n=1}^{\infty}f_{n}\sum_{l=0}^{n-1}\bar{\square}_{s}^{l}\delta^{(2)}\square_{s}\bar{\square}_{s}^{n-l-1}\delta R
=  \int d^{4}x\sqrt{-\bar{g}}\bar{R}\delta^{(2)}\square_{s}\Zc_{1}\left(\bar{\square}_{s}\right)\delta R\\
\approx &  -64\epsilon^4 T_{\textrm{NL}}\int d\tau d^3x a^2 \Hc \Bigg[\bar{R}\Rc^2\Rc^\prime +  \frac{8}{a^2}\Rc\Rc^\prime\Rc^{\prime\prime}-\frac{8}{a^2}\Rc\nabla\Rc\cdot\nabla\Rc^\prime\Bigg] \,.
\end{aligned}
\label{c6f}
\]
The contributions from this term are (\ref{A4}), (\ref{A8}) and (\ref{A9}) 
\begin{equation}
T_4\supseteq -\frac{8\bar{ R}}{M_p^2}\epsilon^3T_{\textrm{NL}},\, \quad T_8 \supseteq -\frac{16\bar{ R}}{3M_p^2}\epsilon^3T_{\textrm{NL}}\,,\quad  T_9 = \frac{16\bar{ R}}{3M_p^2}\epsilon^3T_{\textrm{NL}} \,. 
\end{equation}
Let us compute the second term 
\begin{equation}
\begin{aligned}
	&\int d^{4}x\sqrt{-\bar{g}}\bar{R}\left[\delta^{(2)}\Fc_{R}\left(\bar{\square}_{s}\right)\Bigg\vert_{\delta\square_{s}\delta\square_{s}}\right]\delta R\\
	= & \int d^{4}x\sqrt{-\bar{g}}\bar{R}\sum_{n=2}^{\infty}f_{n}\sum_{l=0}^{n-2}\sum_{k=0}^{n-2-1}\bar{\square}_{s}^{l}\delta\square_s\bar{\square}_{s}^{k}\delta_s\square\bar{\square}_{s}^{n-l-k-2}\delta R\\
=&  \int d^{4}x\sqrt{-\bar{g}}\bar{R}\delta\square_s\sum_{n=2}^{\infty}f_{n}\sum_{l=0}^{n-2}\sum_{k=0}^{n-2-l}\bar{\square}_{s}^{n-2}\delta\square_s\delta R\\
=& \int d^{4}x\sqrt{-\bar{g}}\bar{R}\delta\square_s\LT\Zc_{2}(\bar{\square}_{s})+\frac{\Fc_{R}^{(\dagger)}\LF\frac{M^2}{\Mc_{s}^{2}}\RF}{(\square-r_{1})}\RT\delta\square_s\delta R\,\\
\approx & \,\,\epsilon^2 \int d\tau d^3x a^4 12\frac{\bar{R}}{\Mc_{s}^2}\frac{\Hc}{a^2}\Zc_2\LF \frac{M^2}{\Mc_s^2}+\frac{\bar{R}}{4\Mc_s^2} \RF\Rc^\prime\delta\square_s\delta R\, \\ 
\approx & -192\Mc_{s}^2\epsilon^4 T_{\textrm{NL}} \int d\tau d^3x a^2 \Hc\Rc^\prime \Bigg[ \frac{2M^2}{\Mc_{s}^2}\Rc^2 +\frac{2}{a^2\Mc_{s}^2}\Rc^{\prime 2} -\frac{2}{a^2\Mc_{s}^2}\nabla\Rc\cdot\nabla\Rc \Bigg] \, .
\end{aligned}
\label{c7f}
\end{equation}
where 
\begin{equation}
\mathcal{Z}_{2}\left(\square_s\right)=  \frac{\mathcal{F}_{R}\left(\square_s\right)-\mathcal{F}_{1}}{\left(\square_s-\frac{M^2}{\Mc_s^2}\right)^{2}}\,.
\end{equation}
The leading contributions from this term are (\ref{A5}) and (\ref{A6}) 
\begin{equation}
T_5\supseteq \frac{4\bar{ R}}{M_p^2}\epsilon^3 T_{\textrm{NL }}\,,\quad T_6\supseteq -\frac{4\bar{ R}}{M_p^2}\epsilon^3T_{\textrm{NL }}\,. 
\end{equation}

Similarly, the following non-local contribution can be written as two parts 
	\[
	\begin{aligned}
	\int d^{4}x \delta\LF\sqrt{-g} R\RF\delta^{(2)}\Fc_{R}\left(\bar{\square}_{s}\right)\bar{R} \approx & \int d^{4}x\sqrt{-\bar{g}}\delta R\left[\delta^{(2)}\Fc_{R}\left(\bar{\square}_{s}\right)\Bigg\vert_{\delta^{(2)}\square_{s}}\right]\bar{R} \\ +&\int d^{4}x\sqrt{-\bar{g}}\delta R\left[\delta^{(2)}\Fc_{R}\left(\bar{\square}_{s}\right)\Bigg\vert_{\delta\square_{s}\delta\square_{s}}\right]\bar{ R}\,. 
	\end{aligned}
	\label{c8f}
	\]
 The first term in (\ref{c8f}) can be simplified as 
 \begin{equation}
 \begin{aligned}
& \int d^{4}x\sqrt{-\bar{g}}\delta R\left[\delta^{(2)}\Fc_{R}\left(\bar{\square}_{s}\right)\Bigg\vert_{\delta^{(2)}\square_{s}}\right]\bar{R}=  \int d^4x\sqrt{-\bar{g}}\delta R \Zc_{1}\LF \square_s \RF \delta^{(2)}\square_s \bar{ R}\,\\ 
 \approx & \int d^4x\sqrt{-\bar{g}}\delta R \Zc_{1}\LF \square_s \RF \delta^{(2)}\square_s \bar{ R}\,\\ 
  \approx &\, -64\epsilon^4 T_{\textrm{NL}}\int d\tau d^3x a^2\Hc\Rc^{\prime}\LT 4M^2\Rc^2-\Hc\epsilon\frac{16}{a^2}\Rc\Rc^\prime\RT \,, 
 \end{aligned}
 \label{c8f1}
 \end{equation}
 The bi-spectrum contribution from this term is (\ref{A2})
 \begin{equation}
 T_2  \supseteq -\frac{32\bar{ R}}{3M_p^2}\epsilon^4 T_{NL} \,. 
 \end{equation}
Now the second term in (\ref{c8f}) can be worked out as
		\[
	\begin{aligned}
&	\int d^{4}x\sqrt{-\bar{g}}\delta R\left[\delta^{(2)}\Fc_{R}\left(\bar{\square}_{s}\right)\Bigg\vert_{\delta\square_{s}\delta\square_{s}}\right]\bar{ R}=  \int d\tau d^3x a^4\delta R\delta\square_s\Zc_{2}(\bar{\square}_{s})\delta\square_s\bar{R}\\
	\approx &  \, 4\epsilon T_{\textrm{NL}} \int d\tau d^3x a^2 \LF -2\epsilon\Rc \RF \delta\square_s \Hc \Rc^\prime\,, \\
	\approx &\, 128\epsilon^4T_{\textrm{NL}} \Mc_{s}^2 \int d\tau d^3x a^4 \Rc\Bigg[ \frac{1}{a^2}\Hc\Bigg( \frac{\bar{ R}}{4\Mc_{s}^2}\Rc\Rc^\prime +\frac{2}{\Mc_{s}^2a^2}\Rc^\prime\Rc^{\prime\prime}-\frac{2}{\Mc_{s}^2a^2}\nabla\Rc\cdot\nabla\Rc^{\prime} \Bigg)\Bigg]\,.
	\end{aligned}
	\label{c8f2}
	\]
The leading contributions from this term (\ref{A4}), (\ref{A8}) and (\ref{A9}) 
\begin{equation}
T_4  \supseteq -\frac{8\bar{ R}}{M_p^2}\epsilon^3 T_{NL}\,, \quad T_8 \supseteq -\frac{8\bar{ R}}{3M_p^2}\epsilon^3 T_{NL}\,, \quad T_9  \supseteq \frac{8\bar{ R}}{3M_p^2}\epsilon^3 T_{NL}\,. 
\end{equation}
	
The terms involving 3rd variation of the form-factor can be split into 4 different type of terms as below 
	\[
	\begin{aligned}
		&\int d^{4}x\sqrt{-\bar{g}}\bar{R}\delta^{(3)}\Fc_{R}\left({\square}_{s}\right)\bar{R}\\
		= & \int d^{4}x\sqrt{-\bar{g}}\bar{R}\left[\delta^{(3)}\Fc_{R}\left({\square}_{s}\right)\Bigg\vert_{\delta^{(3)}\square_{s}}\right]\bar{R} +\int d^{4}x\sqrt{-\bar{g}}\bar{R}\left[\delta^{(3)}\Fc_{R}\left({\square}_{s}\right)\Bigg\vert_{\delta^{(2)}\square_{s}\delta\square_{s}}\right]\bar{R}\\
	 +&\int d^{4}x\sqrt{-\bar{g}}\bar{R}\left[\delta^{(3)}\Fc_{R}\left({\square}_{s}\right)\Bigg\vert_{\delta\square_{s}\delta^{(2)}\square_{s}}\right]\bar{R}+\int d^{4}x\sqrt{-\bar{g}}\bar{R}\left[\delta^{(3)}\Fc_{R}\left({\square}_{s}\right)\Bigg\vert_{\delta\square_{s}\delta\square_{s}\delta\square_{s}}\right]\bar{R}
	\end{aligned}
	\label{c9f}
	\]
	Calculating the first term we obtain 
	\begin{equation}
	\int d^{4}x\sqrt{-\bar{g}}\bar{R}\sum_{n=0}^{\infty}\sum\bar{\square}_{s}^{l}\delta^{(3)}\square_s\bar{\square}_{s}^{(n-l-1)}\bar{R}=\int d^{4}x\sqrt{-g}\bar{R}\Zc_{1}\LF\bar{\square}_{s}\RF\delta^{(3)}\square_s\bar{R}=0\,, 
	\label{c9f1}
	\end{equation}
	which follows from (\ref{FD1zero}). 
	Now calculating the second term in (\ref{c9f}) we get
	\begin{equation}
	\begin{aligned}&\int  d^{4}x\sqrt{-\bar{g}}\bar{R}\left[\delta^{(2)}\Fc_{R}\left(\bar{\square}_{s}\right)\Bigg\vert_{\delta^{(2)}\square_{s}\delta\square_{s}}\right]\bar{R}=  \int d^{4}x\sqrt{-g}\bar{R}\delta^{(2)}\square_s\Zc_{2}(\square_s)\delta\square_s\bar{R}\\
	\approx &\, 64\epsilon^4 \Mc_{s}^2 T_{\textrm{NL}} \int d\tau d^3x a^4\Bigg[ \frac{1}{a^2}\Hc\Bigg(\frac{4M^2}{\Mc_{s}^2}\Rc^2\Rc^\prime-\frac{8}{a^2\Mc_{s}^2}\Rc\nabla\Rc\cdot\nabla\Rc^\prime+\frac{8}{a^2\Mc_{s}^2}\Rc\Rc^\prime\Rc^{\prime\prime} \Bigg)\Bigg]\,.\end{aligned}
	\label{c9f2}
	\end{equation}
	The leading contributions from (\ref{c9f2}) are (\ref{A8}) and (\ref{A9})
	\begin{equation}
	T_8 \supseteq \frac{16\bar{ R}}{3M_p^2}\epsilon^3 T_{\textrm{NL}}\,,\quad T_9 \supseteq -\frac{16\bar{ R}}{3M_p^2}\epsilon^3 T_{\textrm{NL}}\, .
	\end{equation}
	We neglect the first interaction term $\Rc\Rc\Rc^\prime$ in (\ref{c9f2}) in comparison with the first term in (\ref{c8f2})  within the slow-roll approximation. 
	
	Let us now consider the third term in (\ref{c9f}) and evaluate it as 
	\begin{equation}
	\begin{aligned}&\int d^{4}x\sqrt{-g}R\sum_{n=0}^{\infty}f_{n}\sum_{a+b+c=n-2}\square^{a}\delta\square\square^{b}\delta^{(2)}\square\square^{c}R\\
		= & \,\epsilon^3\int d^{4}x\sqrt{-g}\bar{R}\delta\square_s\Zc_{2}(\bar{\square}_{s})\delta^{(2)}\square_s\bar{R}
	\approx  \, 192 T_{\textrm{NL}}\epsilon^5\int d\tau d^3x 16\Hc^2 \Rc\Rc^{\prime 2} \,.\end{aligned}
	\end{equation}
	The leading contribution from this term is (\ref{A2}) 
	\begin{equation}
	T_2  \supseteq \frac{32\bar{ R}}{M_p^2} \epsilon^4 T_{\textrm{NL}} \, .
	\end{equation}

Now expanding the last term in (\ref{c9f}) we obtain 
	\begin{equation}
	\begin{aligned} & \int d^4{x}\sqrt{-\bar{g}}\bar{R}\sum_{n=3}^{\infty}f_{n}\sum_{l=0}^{n-3}\sum_{k=0}^{n-l-3}\sum_{m=0}^{n-k-l-3}\square_s^{l}\delta\square_s\square_s^{k}\delta\square_s\square_s^{m}\delta\square_s\square_s^{n-l-k-m-3}\bar{R}= \\
	& \int d^4{x}\sqrt{-\bar{g}}\bar{R}\sum_{n=3}^{\infty}f_{n}\sum_{l=0}^{n-3}\sum_{k=0}^{n-l-3}\sum_{m=0}^{n-k-l-3}\LF \frac{M^2}{\Mc_{s}^2} \RF^{n-k-m-3}\delta\square_s\square_s^{k}\delta\square_s\square_s^{m}\delta\square_s \bar{R}\,\approx\\
	&  -48\tilde{T}_{\textrm{NL}} \frac{64\Mc_{s}^2}{\bar{ R}}\epsilon^5  \int d^3xd\tau\Hc^2\Rc^\prime \Bigg(\frac{\bar{ R}}{2\Mc_{s}^2}\Rc\Rc^\prime -\frac{2}{a^2\Mc_{s}^2}\nabla\Rc\cdot\nabla\Rc^\prime\Bigg) \,, 
	\end{aligned}
	\label{lterm}
	\end{equation}
	where 
	\begin{equation}
	\tilde{T}_{\textrm{NL}} =  \LT \frac{\bar{ R}}{4\Mc_{s}^2}\Fc_{R}^{(\dagger)}\LF \frac{\bar{ R}}{4\Mc_{s}^2}\RF+ T_{\textrm{NL}} \RT
	\end{equation}
	The contributions from this term are (\ref{A2}) and (\ref{A7})
	\begin{equation}
	T_2 \supseteq -\frac{16\bar{ R}}{M_p^2}\epsilon^4 \tilde{T}_{\textrm{NL}}\,,\quad  T_7 \supseteq \frac{16\bar{ R}}{3M_p^2} \epsilon^4 \tilde{T}_{\textrm{NL}}\,. 
	\end{equation}
	In the above derivation we have used the following result 
	\begin{equation}
		\sum _{l=0}^{n-3} \sum _{k=0}^{-l+n-3} \sum _{m=0}^{-k-l+n-3} x^{n-k-m-3} y^{k+m} =\frac{(n-2)(x^{n}-y^n)-n y x^{n-1}+n x y^{n-1}}{ (x-y)^3} \,.  
	\end{equation}
	
	Now we collect all the terms above and calculate the bi-spectrum contribution from each type of interaction substituting the curvature perturbation evaluated in the adiabatic vacuum  initial state (\ref{curpert}) in (\ref{3-point-f}). 
	To illustrate, in the rest of the computations here, we calculate the 3-point correlation contribution from the interaction $\Rc\nabla\Rc\cdot\nabla \Rc$ of (\ref{c1f}) using (\ref{curpert}) and (\ref{3-point-f})
	\begin{equation} 
	\begin{aligned}
	-&i2^5\pi^7 \delta^3\LF \mathbf{k_1}+\mathbf{k_2}+\mathbf{k_{3}} \RF  \frac{1}{\prod_i \LF 2k_i^3 \RF} \frac{H^6}{2^8\pi^4\epsilon^4} \\ \times& 64\epsilon\LF \mathbf{k_{1}}\cdot \mathbf{k_{2}} \RF \int_{-\infty}^{\tau_e} d\tau \frac{1}{\tau^2}\LF 1-ik_1\tau \RF\LF 1-ik_2\tau \RF\LF 1-ik_3\tau\RF e^{iK\tau}
	\end{aligned}
	\end{equation}

		\begin{enumerate}
		\item Calculating the bi-spectrum contribution due to the interaction $\Rc\nabla\Rc\cdot\nabla \Rc$, we get 
			\begin{equation}
		2^5\pi^7\delta^{3}\LF \mathbf{k_1}+\mathbf{k_2}+\mathbf{k_3}\RF \frac{1}{\prod_i k_i^3} \Pc_{\Rc}^{2}T_12\LF \boldsymbol{k}_1\cdot \boldsymbol{k}_2 \RF  \LT -K+\frac{\sum_{i>j} k_ik_j}{K} +\frac{k_1k_2k_3}{K^2} \RT +\text{perms}\footnote{where the additional terms are obtained by cycling permutations of momenta
.}\,.
		\label{A1}
		\end{equation}
				\item Calculating the bi-spectrum contribution due to the interaction $\Rc\Rc^{\prime 2}$, we get 
		\begin{equation}
		2^5\pi^7 \delta^{3}\LF \mathbf{k_1}+\mathbf{k_2}+\mathbf{k_3}\RF \frac{1}{\prod_i k_i^3} \Pc_{\Rc}^{ 2} T_2 \LT \frac{2k_1^2k_2^2}{K} +\frac{2k_1^2k_2^2k_3}{K^2} \RT +\text{perms}\,.
		\label{A2}
		\end{equation}	
	\item Calculating the bi-spectrum contribution due to interaction $\Rc^3$, we obtain 
		\begin{equation}
		2^5\pi^7 \delta^{3}\LF \mathbf{k_1}+\mathbf{k_2}+\mathbf{k_3}\RF \frac{1}{\prod_i k_i^3} \Pc_{\Rc}^{2}  T_3
		\LT - \frac{K^3}{3} +2Kk_1k_2+\frac{k_1k_2k_3}{3} \RT +\text{perms}\,. 
		\label{A3}
		\end{equation}
			\item Calculating the bi-spectrum contribution due to the interaction $\Rc\Rc\Rc^\prime$, we obtain
		\begin{equation}
		2^5\pi^7 \delta^{3}\LF \mathbf{k_1}+\mathbf{k_2}+\mathbf{k_3}\RF \frac{1}{\prod_i k_i^3} \Pc_{\Rc}^{2} T_4 k_3^2\LT -\frac{4K}{3}-\frac{2k_1k_2}{K} -\frac{2k_1+2k_2}{3} \RT +\text{perms}\,.
		\label{A4}
		\end{equation}
			\item Calculation of the bi-spectrum contribution due to the interaction $\nabla\Rc\cdot\nabla\Rc\Rc^\prime$ gives 
		\begin{equation}
		2^5\pi^7 \delta^{3}\LF \mathbf{k_1}+\mathbf{k_2}+\mathbf{k_3}\RF \frac{1}{\prod_i k_i^3} \Pc_{\Rc}^{ 2} T_5 2\LF \boldsymbol{k}_1\cdot \boldsymbol{k}_2 \RF k_3^2\LT \frac{2}{K}+\frac{2k_1+2k_2}{K^2}+\frac{4k_1k_2}{K^3} \RT +\text{perms}\,. 
		\label{A5}
		\end{equation}
		\item Calculating the bi-spectrum contribution due to the interaction $\Rc^{\prime 3}$, we obtain
		\begin{equation}
		2^5\pi^7 \delta^{3}\LF \mathbf{k_1}+\mathbf{k_2}+\mathbf{k_3}\RF\frac{1}{\prod_i k_i^3} \Pc_{\Rc}^{2} T_6 \frac{4k_1^2k_2^2k_3^2}{K^3}+\text{perms}\,. 
		\label{A6}
		\end{equation}
		\item Calculating the bi-spectrum contribution due to the interaction  $\Rc^\prime \nabla\Rc\cdot\nabla\Rc^\prime$, we get
		\begin{equation}
		2^5\pi^7 \delta^{3}\LF \mathbf{k_1}+\mathbf{k_2}+\mathbf{k_3}\RF\frac{1}{\prod_i k_i^3} \Pc_{\Rc}^{2} T_7\LF \boldsymbol{k}_2\cdot \boldsymbol{k}_3 \RF k_1^2k_3^2\LT -\frac{2}{K^3}-\frac{6k_2}{K^4} \RT +\text{perms}\,. 
		\label{A7}
		\end{equation}
		\item Calculating the bi-spectrum contribution due to the interaction $\Rc\Rc^{\prime}\Rc^{\prime\prime}$, we get 
		\begin{equation}
	2^5\pi^7\delta^{3}\LF \mathbf{k_1}+\mathbf{k_2}+\mathbf{k_3}\RF\frac{1}{\prod_i k_i^3} \Pc_{\Rc}^{ 2} 	T_8 k_2^2k_3^2\LT \frac{1}{K}+\frac{k_1-k_3}{K^2}-\frac{2k_1k_3}{K^3} \RT+\text{perms}\,. 
			\label{A8}
		\end{equation}		
				\item Calculating the bi-spectrum contribution due to the interaction  $\Rc\nabla\Rc\cdot\nabla\Rc^\prime$, we get
		\begin{equation}
		2^5\pi^7 \delta^{3}\LF \mathbf{k_1}+\mathbf{k_2}+\mathbf{k_3}\RF\frac{1}{\prod_i k_i^3} \Pc_{\Rc}^{ 2} T_9\LF \boldsymbol{k}_1\cdot \boldsymbol{k}_2 \RF  k_2^2\LT \frac{1}{K}+\frac{k_2+k_3}{K^2}+\frac{2k_2k_3}{K^3} \RT+\text{perms}\,. 
		\label{A9}
		\end{equation}
	\end{enumerate} 
where
\begin{equation}
\begin{aligned}
	T_1 & = -2\epsilon-\frac{3\epsilon^2}{4},\quad 
	T_2 =  \LT 2\epsilon +\frac{3\epsilon^2}{4}+ \frac{16\bar{ R}}{3M_p^2}\epsilon^4 T_{\textrm{NL}} + \frac{4\bar{ R}^2}{M_p^2\Mc_{s}^2}\epsilon^4\Fc_{R}^{(\dagger)}\LF \frac{\bar{ R}}{4\Mc_{s}^2}\RF \RT\Bigg\vert_{K=aH},\\
T_3 & = -\frac{\epsilon^2}{2} ,\quad 
T_4  =  -T_{\textrm{NL}} \frac{32\bar{ R}}{M_p^2}\epsilon^3\Bigg\vert_{K=aH},\quad 
T_5 = T_{\textrm{NL}}\frac{2\bar{ R}}{M_p^2}\epsilon^3\Bigg\vert_{K=aH} ,\quad 
T_6  = -T_{\textrm{NL}}\frac{2\bar{ R}}{M_p^2}\epsilon^3 \Bigg\vert_{K=aH},\\ 
T_7 & = \LT\frac{16\bar{ R}}{3M_p^2}\epsilon^4 T_{\textrm{NL}}+\frac{4\bar{ R}^2}{3M_p^2\Mc_{s}^2}\epsilon^4\Fc_{R}^{(\dagger)}\LF \frac{\bar{ R}}{4\Mc_{s}^2}\RF\RT \Bigg\vert_{K=aH},\quad 
T_8 = 0,\quad 
T_9  = 0 \,.  
\end{aligned}
\label{Tterms}
\end{equation}

We have used the following integrals in the calculation of 3-point functions
\begin{equation}
\begin{aligned}
	&Re\LT  -i\int^{\tau_e}_{-\infty}  e^{iK\tau} \RT  = \frac{1}{K},\quad
	Re\LT  -i\int^{\tau_e}_{-\infty}  \tau^ne^{iK\tau} \RT  = \frac{\pd^n}{\pd K^n}\frac{1}{K},\\ 
	I_1=&\int_{-\infty}^{\tau_e} d\tau \frac{1}{\tau}e^{iK\tau}=i\pi+\mathrm{Ei}(iK\tau_e),\quad
	I_n  = \int_{-\infty}^{\tau_e} d\tau \frac{1}{\tau^n}e^{iK\tau},\quad
I_{n+1}  = -\frac{1}{n}\LF \frac{1}{\tau_e}\RF^n e^{iK\tau_e}+\frac{iK}{n}I_n \,. 
\end{aligned}
\end{equation}
Here $\mathrm{Ei}(z)$ is the integral exponential function.
The integral has to be evaluated in the limit $\tau_e\to 0$ while convergence at large $\tau$ is made possible by the oscillatory behaviour at $\tau \to -\infty $. Some of the integrals do diverge in the limit $\tau_e \to 0$ where we must follow the guidance well explained in \cite{Pajer:2016ieg} which prescribes us to evaluate integrals up to the conformal times corresponding to the Hubble radius crossing scale, such that $K\sim aH\sim -\frac{O(1)}{\tau_e}$. For the purposes of practical computations one fixes $K\tau_e=-1$.



\providecommand{\href}[2]{#2}\begingroup\raggedright\endgroup
\end{document}